\documentclass[a4paper]{article}
\usepackage[utf8]{inputenc}

\usepackage{hyperref}
\usepackage{booktabs}
\usepackage{subfigure, dsfont, multirow, array}
\usepackage{graphicx}
\usepackage[a4paper, margin=0.8in]{geometry}
\usepackage[11pt]{moresize}

\title{Predicting Smartphone Battery Life based on Comprehensive and Real-time Usage Data}
\author{
	Huoran Li, {\small Peking University, lihuoran@pku.edu.cn} \\
	Xuanzhe Liu, {\small Peking University, liuxuanzhe@pku.edu.cn} \\
	Qiaozhu Mei, {\small University of Michigan, qmei@umich.edu}	
}

\begin{document}

\maketitle

\begin{abstract}
Smartphones and smartphone apps have undergone an explosive growth in the past decade. However, smartphone battery technology hasn't been able to keep pace with the rapid growth of the capacity and the functionality of smartphones and apps. As a result, battery has always been a bottleneck of a user's daily experience of smartphones. An accurate estimation of the remaining battery life could tremendously help the user to schedule their activities and use their smartphones more efficiently. Existing studies on battery life prediction have been primitive due to the lack of real-world smartphone usage data at scale. This paper presents a novel method that uses the state-of-the-art machine learning models for battery life prediction, based on comprehensive and real-time usage traces collected from smartphones. The proposed method is the first that identifies and addresses the severe data missing problem in this context, using a principled statistical metric called the concordance index. The method is evaluated using a dataset collected from 51 users for 21 months, which covers comprehensive and fine-grained smartphone usage traces including system status, sensor indicators, system events, and app status. We find that the remaining battery life of a smartphone can be accurately predicted based on how the user uses the device at the real-time, in the current session, and in history. The machine learning models successfully identify predictive features for battery life and their applicable scenarios.
\end{abstract}


\section{Introduction}

How many times were you expecting an important conference call but found only 20\% battery left on your iPhone? How many times were you in the middle of a city tour but felt nervous about turning on the direction service, with the fear of not having enough battery for the rest of the day? Should I find a restaurant to stop and recharge, or should I do it after that mountain hike? What if I tell you that your battery can still last for one hour? 

Ever since the introduction of the first iPhone in 2007, we have been witnessing the tremendous growth of smartphones and smartphone apps. Today, smartphones have been an indispensable component of our daily lives. The rich functionalities of a smartphone have expanded its role way beyond a communication tool and towards a pervasive computing device. 
As of 2017, there are over a hundred popular models of smartphone devices on the market, all enpowered with multiple core processors, large screens, RAMs with comparable sizes to personal computers, hundreds of Gigabytes of ROMs, and a variety of sensors\footnote{\url{https://oneplus.net/5}}. Meanwhile, there are more than 2,200,000 and 2,800,000 apps on marketplaces such as Apple Appstore and Google Play, respectively\footnote{\url{https://www.statista.com/statistics/276623/number-of-apps-available-in-leading-app-stores/}}. The co-growth of hardware and software makes smartphones more powerful than it has ever been. 

Unfortunately, the improvement of battery technology has not been able to keep pace with the rapid growth of the devices and apps. For most smartphones, the battery could last for at most one day, and the battery life drops rapidly when the usage is heavier. Short battery life has always been the bottleneck of a user's daily experience of smartphones. As a result, users often need to bring alternative power sources such as spare batteries or power banks when it is inconvenient to charge, or their smartphone activities have to be compromised due to battery limits. It is by all means critical to explore ways to ease the battery-caused pains of smartphone users. 

In literature, much effort has been done to approach this problem, such as to analyze and reduce energy consumption of a specific hardware component~\cite{balasubramanian2009energy, puustinen2011effect, rosen2015revisiting, shen2015enhancing}, to restrict unnecessary resource usage~\cite{he2015optimizing, chen2015smartphone, draa2015application}, and so on. Most of them attempt to extend the battery life by enhancing the efficiency of energy consumption.  
In practice, such approaches to extended battery life (e.g., the ``energy saving mode'') are usually at the risk of reducing the functionalities and compromising the user experience. Instead, an intriguing question to ask alternatively is how to optimize the experience of users under the battery capacity limits. 

Along this line, the very first and critical step is to let the user know how much longer the battery would still last before it has to be recharged, a metric that can be described as the \textit{remaining lifetime of the battery} or simply the \textit{battery life}. If the battery life can be accurately estimated at the real time, the user is able to plan their smartphone usage smartly, either to carry on the current activities, or to take actions to save the energy for an important activity scheduled. However, due to the complexity of the hardware and software of smartphones, predicting the remaining battery life in real time is not a trivial task. Indeed, it is almost infeasible to identify the energy drain of every currently running app and every hardware component. The expected battery life time, however, doesn't just depend on the current or past energy consumption status, but is also highly influenced by the various activities the user plans to perform next, and these activities themselves are hard to predict~\cite{li2015characterizing, falaki2010diversity}. 

Existing efforts on predicting battery life at real time are far from meeting these challenges. In fact, most of them appear to be too primitive and not practical due to the following reasons. First, the lack of fine-grained data has been a big obstacle to almost every existing study. Most data sets used in literature are collected from lab experiment, so they are short-term and usually do not represent the usage of smartphones in reality. Second, these data sets usually only collect a few surface data attributes at a coarse granularity (e.g., readings of battery levels at every a few minutes) instead of in-depth and fine-grained signals (e.g., sensor readings or screen statuses at seconds level). Due to the lack of rich and longitudinal signals, most existing work is only able to explore simple predictors rather than the much more powerful machine learning models. Moreover, 
none of the existing studies has been able to identify and handle a classical and severe data missing problem in this prediction task. To be specific, it is critical to distinguish the cases where a user charges their phone before the battery dies from the cases where the battery actually dies. Failing to treat the two cases appropriately may result in significant biases and compromised accuracy in the prediction task. 

We present the first exploration of the modern ``AI + Big Data'' approach to battery life prediction, based on state-of-the-art machine learning algorithms and comprehensive, fine-grained signals collected from real-world uses of smartphones. Our analysis is based on a data set~\cite{mirsky2016sherlock} that collected fine-grained traces of system statuses, system events, sensor readings, and app statuses from the daily usage of 51 smartphone users in 21 months. The data set contains \textbf{every software and hardware sensor} that can be obtained from a Samsung Galaxy S5 smartphone without root privileges. To the best of our knowledge, this is the most fine-grained and comprehensive data set that has ever been used for this purpose. This is also the first work that formally identifies and handles the data missing problem in the battery life prediction, and the first that systematically evaluates advanced machine learning models for this task. 

In summary, we make the following contributions in this paper:

\begin{itemize}
    \item We conduct a descriptive analysis of how users consume the energy of their smartphones in daily uses. We find that almost half of usage sessions begin with not a full level of battery, and about 40\% of the usage sessions are exposed to the risk of running out of power.
    \item We present a systematic evaluation of how well the system statuses and user behaviors can join forces through state-of-the-art machine learning models to predict the battery life. Surprisingly, we find that the battery level and system statuses at the present are not sufficiently predictive for the expected battery life. Instead, the battery's discharging history, past system statuses in the session, and the user's historical usage patterns may significantly enhance the prediction accuracy. Our model successfully reduces the prediction error by more than 30 minutes on average, comparing to the baseline. 
    \item We formally identify the \textbf{data missing} problem (i.e., users will not always consume all the energy) in battery life prediction and use a classical statistical metric, the \textbf{concordance index}, to address the issue.
    \item We present insights on the scenarios in which different factors are more predictive for the battery life. The experiments show that system statuses in the past perform the best when the energy consuming rate is stable, while user behaviors are more predictive in the opposite situation.
\end{itemize}

The results of our analysis can be easily applied to the real world. The prediction model allows a user to query their battery life at anytime they want, and it predicts how long the battery is expected to last before it needs to be recharged. Although our data set was collected from Samsung Galaxy S5 smartphones, we intentionally avoided using any device-specific feature. Therefore, the method is generalizable to other devices, as long as the devices have the mechanism to collect similar signals. In a long shot, our method and conclusions are not limited to smartphones, but can be applied to predicting the battery life for other smart devices, such as tablets, smart watches, and even electronic vehicles. 

The rest of the paper is organized as follows. We first present the related work in Section 2, followed by an introduction of the data set in Section 3. We formally define our problem in Section 4 and present a descriptive analysis of smartphone battery usage in Section 5. The setup and the results of our experiments are presented in Section 6 and Section 7 respectively. Finally, we discuss the potential limitation of the work in Section 8 and conclude the study in Section 9.


\section{Related Work}

Battery is arguably one of the most important components on mobile devices such as smartphones and wearables, the capacity of which significantly affects user experience. A large body of literature has been trying to address problems related to battery optimization. Generally, existing studies can be partitioned into three categories: power modeling, power saving, and battery life prediction.

One common step of smartphone battery optimization is to understand where and how the energy is consumed on a smartphone. However, it is not trivial to obtain such information directly from the system. Therefore, many studies focus on how to measure the energy drain of every single component of a smartphone, more specifically, through the so-called ``power modeling'' process. Currently, there are two major types of power models. The first one refers to the \textit{utilization-based models}. Power models of this kind are based on the assumption that the power state of a smartphone is relevant to the utilization of its hardware components. For example, Shye et al.~\cite{shye2009into} presented a regression-based power model that uses high-level system measurements to estimate smartphone power consumption. They took various statuses of the hardware (e.g., CPU frequency, RAM, and so on) as input and the power consuming rate as output, then used a linear regression model to derive their relationship. Dong et al.~\cite{dong2011self} implemented a power model which could measure battery consumption at a high rate. This model requires only the battery interface provided by the system, so it does not need any external power measurements. Zhang et al.~\cite{zhang2010accurate} developed a system-level power model that takes into account CPU, LCD, GPS, network modules, and audio components. They also introduced an automated technique to construct utilization-based power models. Min et al.~\cite{min2015sandra} claimed that the battery consumption of continuous sensing apps (CSA) will be strongly influenced by users' mobility (for example, walking or running). They set up a series of experiments to validate their hypotheses. Some other power models try to measure the energy usage of apps. For example, Min et al.~\cite{min2015powerforecaster} designed the PowerForecaster, which uses the  trace-driven emulation to provide an instant and personalized power estimation of a sensing app at the pre-installation time. Dong et al.~\cite{dong2014rethink} regarded that energy accounting of apps should be formulated as a cooperative game problem. Following this principle, they provided a Shapley-value-based energy accounting method to measure apps' energy consumption.

The second category of power models uses Finite-State-Machines (FSMs) to characterize the power consumption of wireless network components. For wireless network components, their power states are not only determined by their utilization, but also related to other, more detailed internal states. Therefore, FSMs are required to capture the power consumption behavior of these components. For example, Pathak et al.~\cite{pathak2011fine} observed that some system calls that do not imply utilization can still change power states. Based on this finding, they proposed an FSM that takes system calls as triggers to model the power consumption behavior of smartphones. The similar principle was also adopted by another energy consumption estimation tool called WattsOn~\cite{mittal2012empowering}. WattsOn introduced a protocol called the Power Save Mode~\cite{manweiler2011avoiding} to capture the power consumption behavior of Wi-Fi modules.

Power modeling provides a foundation for power saving research. In a nutshell, the core idea of existing power-saving strategies is to restrict unnecessary hardware and software usage. For example, He et al.~\cite{he2015optimizing} proposed the Dynamic Resolution Scaling (DRS), which could dynamically adjust the user-interface resolution based on the viewing distance. In this way, DRS could reduce the power consumption caused by high-density displays without compromising user experience. Chen et al.~\cite{chen2015smartphone} presented a system called HUSH to identify unnecessary background activities during screen-off sessions. By killing these activities, HUSH could save screen-off energy by 15.7\% on average. Li et al.~\cite{li2014making} developed an method that can automatically
rewrite web applications so as to generate
more energy-efficient web pages for OLED screens. The key idea of this method is adjusting the background color of web pages to avoid inefficient display along with the possible energy drain. In general, all power-saving strategies need to find a balance between reduced energy consumption and compromised utility and user experience. 

As an alternative strategy, battery life prediction has also drawn much attention in recent years. An accurate battery life prediction model can provide users informative suggestions to opitmize the app usage under battery limits. Zhao et al.~\cite{zhao2011system} proposed a system context-aware approach for battery life prediction. The authors used a linear-regression model to calculate the device's discharging rate based on system contexts (CPU, screen, I/O, etc.), and took this discharging rate as an estimation of future discharging rate. In this way, the remaining battery life can be calculated. Kang et al.~\cite{kang2011personalized} proposed another method from the user's perspective. They divided users' usage behavior into some possible usage states according to apps' functionalities, e.g., browsing a webpage, sending a text message, making a voice call, etc. They simply assumed that the discharging rate under each state is nearly constant. Based on such an assumption, they used basic statistic methods to calculate the discharging rate and the percentage of time under each state and then used these two vectors to predict the battery life. Kim et al.~\cite{kim2016accurate} proposed a framework that can predict the available battery time for a specific app. For example, they aimed to tell users the remaining battery life if the users keep interacting with the devices such as watching videos or listening to music. The state division in this study is based on hardware status rather than the actual usage behaviors. The framework first detects the discharging rate under each state, and then estimates the percentage of battery time under each state while using a specific app.

All these studies mentioned above made meaningful attempts for battery life prediction. The success is however limited by the data sets used in these studies. On the one hand, the data used in most existing studies were collected from either controlled lab experiments or small-scale field studies, usually in a short period of time. In addition, these data sets usually only collect limited types of smartphone usage and system status signals, at a coarse granularity. For example, some previous studies only adopt CPU frequency, Wi-Fi status (turned on or turned off), and screen brightness to predict battery life. These signals are not enough to form a comprehensive representation of the system status, so the prediction is unlikely to be accurate. 
With limited usable attributes, most existing studies can only construct simple statistical predictors and are unable to explore the much more powerful machine learning models. 
Besides the limitation of data, some existing studies also rely on potentially problematic assumptions. For example, the assumption that ``the discharging rate under a specific usage state is constant'' may not hold in practice, not to mention that the usage state itself is never a constant. 

Our work also addresses battery life prediction, but distinguishes from the existing studies in two folds. First, our study is built upon a comprehensive, longitudinal, and fine-grained data set, which collects the real-world smartphone usage traces of 51 users in a 21-month time span. Such a data set allows us to consider the readings from more than one hundred types of sensors, detailed information of system-level events, and various types of user activities. These signals were sampled with a frequency as high as every 5 seconds. Such a data set breaks down the battery usage at very fine granularity, supports the construction of the state-of-the-art machine learning predictors, and enables us to derive much more comprehensive insights. 

In addition, our approach does not rely on arbitrary assumptions, nor does it impose any restriction to the actual queries (for example, some existing work only allows a user to query the predictor when the battery begins to discharge). In this way, our method is more practical and can be applied to most of mobile devices. Comparing to existing work that adopted only basic statistics and linear predictors, we employ a portfolio of state-of-the-art machine learning models, which are more likely to capture complex, non-linear correlations among the signals. Rather than using only the present system statuses or user activities, our approach takes into account two new groups of information, i.e., the past statuses and activities within the same discharging session, and the historical usage behavior of a user. We demonstrate that the proposed approach effectively improves the prediction accuracy. Last but not the least, we makes the first step to identify and deal with a severe \textit{data-missing problem} in battery life prediction, by borrowing concepts from the classical ``survival analysis'' literature.


Although our work follows a standard machine learning practice, it also make an interesting contribution to the machine learning community. That is, we first introduce the concept of concordance index from the survival analysis literature to address the data missing problem. Concordance index as a rigorous evaluation metric, may become a standard in prediction tasks of this kind.


\section{The Sherlock Data Set}

This study is based on the Sherlock data set~\cite{mirsky2016sherlock}, which is a multi-year data collection maintained by the BGU Cyber Security Research Center. The incentive of this data set is to provide mobile security researchers with labeled data, but the data set itself could also be used by any project in mobile data analysis area. This data set is essentially a time-series including nearly \textbf{every software and hardware sensor} that can be got from a Samsung Galaxy S5 smartphone without root privileges.

The Sherlock data set is collected through an Android app (named as ``agent'' in the rest of this paper), which was implemented based on Google Funf framework~\cite{aharony2011social}. To collect data, the authors recruited a group of volunteers, then provide each of them with a Samsung Galaxy S5 smartphone which pre-installed the agent. The volunteers are required to use this smartphone as their major device in their daily life, and their usage data will be recorded. 

Basically, the Sherlock data set has two kinds of data according to how the data are collected: ``PUSH'' data and ``PULL'' data. ``PUSH'' data are collected through event-based approaches, such as recording when an SMS arrives or when the screen is turned on. ``PULL'' data are collected periodically, such as periodically sampling the CPU utilization or the device's acceleration. ``PUSH'' data contain six groups: call log, SMS log, screen status, user presence, broadcast intents, and app packages. Most of these groups' content could be identified by their name. ``PULL'' data contain eight groups: Wi-Fi status, Bluetooth status, application status, and other 5 groups of system status and sensors. Except basic information such as the SDK version of the system, the sampling frequency of ``PULL'' data is at least once a minute, while it can be up to once per 5 seconds for some dimensions.

In total, the Sherlock data set contains 553 data fields, and it has already published 21 months data (spanning from January, 2015 to September, 2016) of 51 participants. In conclusion, the Sherlock data set is not only comprehensive, but also longitudinal and fine-grained, which makes it a great material for mobile data analysis. More details about the data set could be found in the referred paper and on its official website\footnote{\url{http://bigdata.ise.bgu.ac.il/sherlock/\#/}}.

\subsection{User Privacy}

To protect the privacy of the volunteers, several concerns were addressed. First, all data are collected under the volunteers' permission. In addition, all the users' identifiers such as SSIDs, cell tower IDs, and MAC addresses have been hashed, and the phone numbers in text messages and call logs are also hashed as well. Each user is labelled as an non-meaningful string. Moreover, instead of reporting the exact geolocation of the volunteers, the Sherlock data set performs a K-Means clustering over the volunteers' positions, and then use the clustering results to represent users' geolocation. In this way, no user privacy will be leaked through the Sherlock data set.

\section{Problem Formulation}

One major goal of our study is how to predict battery life based on system status and user behavior. In this section, we first define related objects and their notations, then describe the problem definition. The summary of notations is listed in Table~\ref{tab:notation}.

\begin{table}[htbp]
\centering
\caption{Summary of Notations.}
\label{tab:notation}
\begin{tabular}{|l|l|}
\hline
\textbf{Notation} & \textbf{Explanation} \\
\hline
$t_{start}$ & The starting time of a session. \\
$t_{end}$ & The ending time of a session. \\
$t_{event}$ & The time when low battery event occurs. \\
$t_{query}$ & The time when user asking the battery life. (Query time) \\
$L$ & The threshold that triggers low battery event. \\
$b(t)$ & The battery level of a given time $t$. \\
$life(t)$ & The remaining battery life starts from a given time $t$. \\
\hline
\end{tabular}
\end{table}

\subsection{Discharging Session and Battery Function}

We define a \textit{discharging session} (\textbf{session} for short) as a duration spanning from the beginning of discharging to the end of discharging. In practice, a session corresponds to a continuous period of using time. For example, the time can range from unplugging the charger in the morning to plugging the adapter back before going to bed. In the following parts of this paper, we take session as the basic unit for further analysis. Formally, we use $[t_{start}, t_{end}]$ to denote a session, where $t_{start}$ and $t_{end}$ represent the beginning time and ending time of the session, respectively. Naturally, $t_{end}$ is always larger than $t_{start}$. 

In addition, we employ a function $b(t)$ to represent the battery level of current smartphone at a given time point $t$. In this case, the beginning and ending battery level of a session can be denoted as $b(t_{start})$ and $b(t_{end})$. It is easy to understand that the battery level never goes up within a discharging session, so $b(t)$ monotonically decreases along with the increasing $t$.

\subsection{Low-Battery Event and Battery Life}
\label{sec:low-battery-event-and-battery-lifetime}

We define a session's \textbf{low-battery event} as the first time when the battery level degrades to a threshold $L$ ($0\leq L\leq 100$). Particularly, the low battery event refers to the ``run-out-of-power'' event when $L$ is zero. The time when a low-battery event happens is denoted as $t_{event}$. It is worth mentioning that a low-battery event may not happen if $b(t_{end})>L$, i.e., the session ends before the battery level reaches the threshold. Therefore, sessions can be labelled as \textbf{``observed''} or \textbf{``unobserved''} according to whether the low battery event can be observed within the session. The threshold $L$ could significantly influence the ratio between observed sessions and unobserved sessions. 

Note that how to handle the ``unobserved'' sessions is critical. Since these sessions do not have output labels, most existing work decided to exclude them from the prediction task. This treatment is risky and introduces a severe data missing problem, which not only introduces biases into the analysis but also makes the measure of success unreliable. Later in the paper, we will introduce a principled method to deal with the unobserved sessions. 

For observed sessions, the battery life is defined as the time duration from a given time $t$ to $t_{event}$. This can be denoted as a function $life(t)$, whose definition is $life(t)=t_{event}-t$. It should be noticed that $life(t)$ is meaningful only when $t$ is prior to $t_{event}$. We need to emphasize that the ``battery life'' defined here is more general than the ``physical'' battery life in common sense. According to our definition, a battery runs out of its ``life'' if the battery level degrades below a threshold $L$ (e.g., 10\% or 20\%), rather than the battery being completely drained. The duration to the state that the battery is completely drained is a special case when $L=0$. Intuitively, it is more straightforward to predict the time when the battery is completely empty (i.e. $L=0$). However, such a case  does not frequently occur in real life as most people won't wait until then to recharge their phones. Predicting when the battery would completely die is not only less meaningful in practice, but also more difficult because of fewer training examples. A good selection of the threshold $L$ balances the difficulty and the usefulness of the prediction task. We will discuss more details in the following part of this paper.


\subsection{A Regression Problem}

Based on the definitions above, our goal is to predict $life(t)$ from a given time $t$. To be more specific, the prediction target is the remaining time till the battery is considered low (less than the given threshold). It is then natural to consider such a prediction task as a regression problem. More specifically, given any set of information of time $t$, we train a regression model that predicts $life(t)$ for the session. The inputs of the regression model contain system status, user's usage behavior, and user's usage history. More details will be discussed in the following sections. 
\section{A Descriptive Analysis}

To demonstrate the representativeness of out data set and to better understand users' battery-usage pattern, we first make a descriptive analysis of our data set in this section. As we have mentioned previously, we take session as the elementary unit during the analysis. Therefore, we first introduce how to generate sessions out of the raw data. 

\subsection{Session Generation}
\label{sec:session-generation}

The battery level data belongs to ``PULL'' data of the Sherlock data set. The data-collection agent reads the battery level of Android OS for approximately every 5 seconds. Each entry of the battery level data contains four fields: 1) user index, 2) timestamp, 3) charging status, and 4) battery percentage. In this case, we can know whether the user's device is charging or discharging along with the battery percentage level at a specific time point. For example, a data entry ``$(0a50e09262, 1426245782, discharge, 54)$'' means that the battery of user ID \textit{0a50e09262} is discharging at the time \textit{1426245782}, and the current battery level is 54\%. 

To generate sessions, we first discard all charging entries, after which we can get a bunch of continuous series of discharging entries. Since the interval between two consecutive entries is only 5 seconds, it is reasonable to trust that the battery keeps discharging between two consecutive discharging entries, unless either of the following two conditions holds:

\begin{enumerate}
    \item The interval between the entries is larger than a large threshold. Such a case could happen as some data are not recorded by the agent. In this case, it cannot be guaranteed that the battery keeps discharging during this period of time. To be more rigorous, we cut one session into two whenever this condition holds. In this work, we set this threshold to be 10 minutes.
    \item The battery level of the latter entry is higher than that of the former one. The explanation of such a phenomenon is that the user replaced a new battery, so the latter entry should be considered as the start of a new session. 
\end{enumerate}

In this way, the original battery-level data could be transformed into some sessions. To make the battery life prediction model reliable and meaningful, we need to further filter out sessions whose lifecycle is too short. For simplicity, we take one hour as the threshold. We believe that sessions shorter than one hour are mostly temporary usage. Their usage pattern is more likely to have a big difference with normal usage, which will affect the prediction's accuracy. Finally, we obtained 37,088 sessions in total. 

\subsection{Distribution of Battery Usage}

\begin{figure}
	\centering
	\begin{center}
		\includegraphics[width=0.5\textwidth]{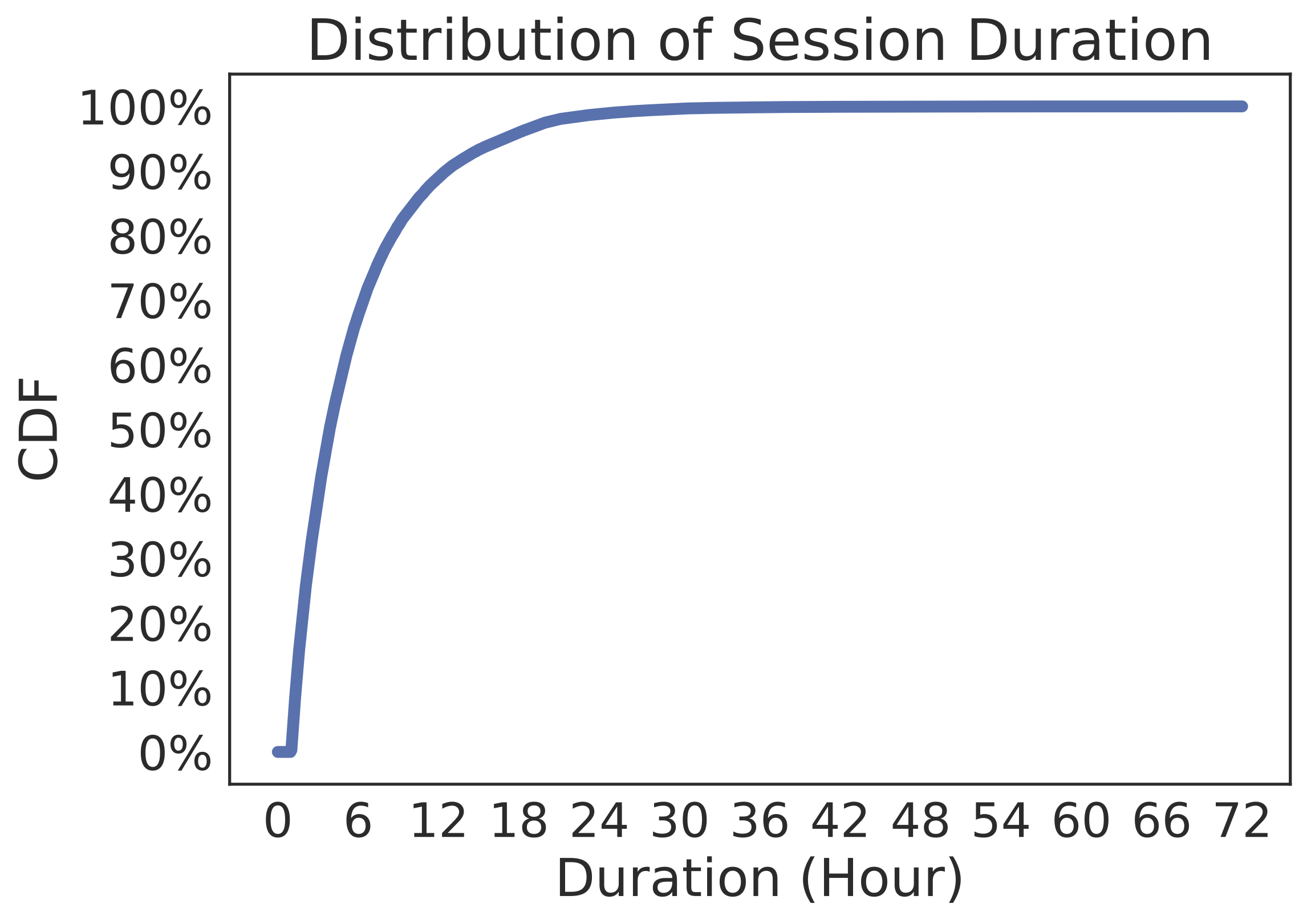}
		\caption[7.5pt]{Cumulative distribution function (CDF) of session duration.}
		\label{fig:cdf-session-duration}
	\end{center}
\end{figure}

We then report the distributions that can be helpful to validate the representativeness of the data and to motivate the following studies. We start with the distribution of the duration time of the sessions, which is demonstrated in Figure~\ref{fig:cdf-session-duration}. The longest duration is about 71 hours, while the shortest duration is about one hour since we have filtered out sessions that are shorter than this threshold. We can also observe that most sessions are very short. More specifically, 90\% of sessions are shorter than 13 hours, while less than 2\% sessions are longer than one day. Indeed, these results are quite consistent with our common experience. 

\begin{figure}
	\centering
	\begin{center}
		\subfigure[Beginning Battery Level\label{fig:cdf-battery-beginning}]
		{\includegraphics[width=0.32\textwidth]{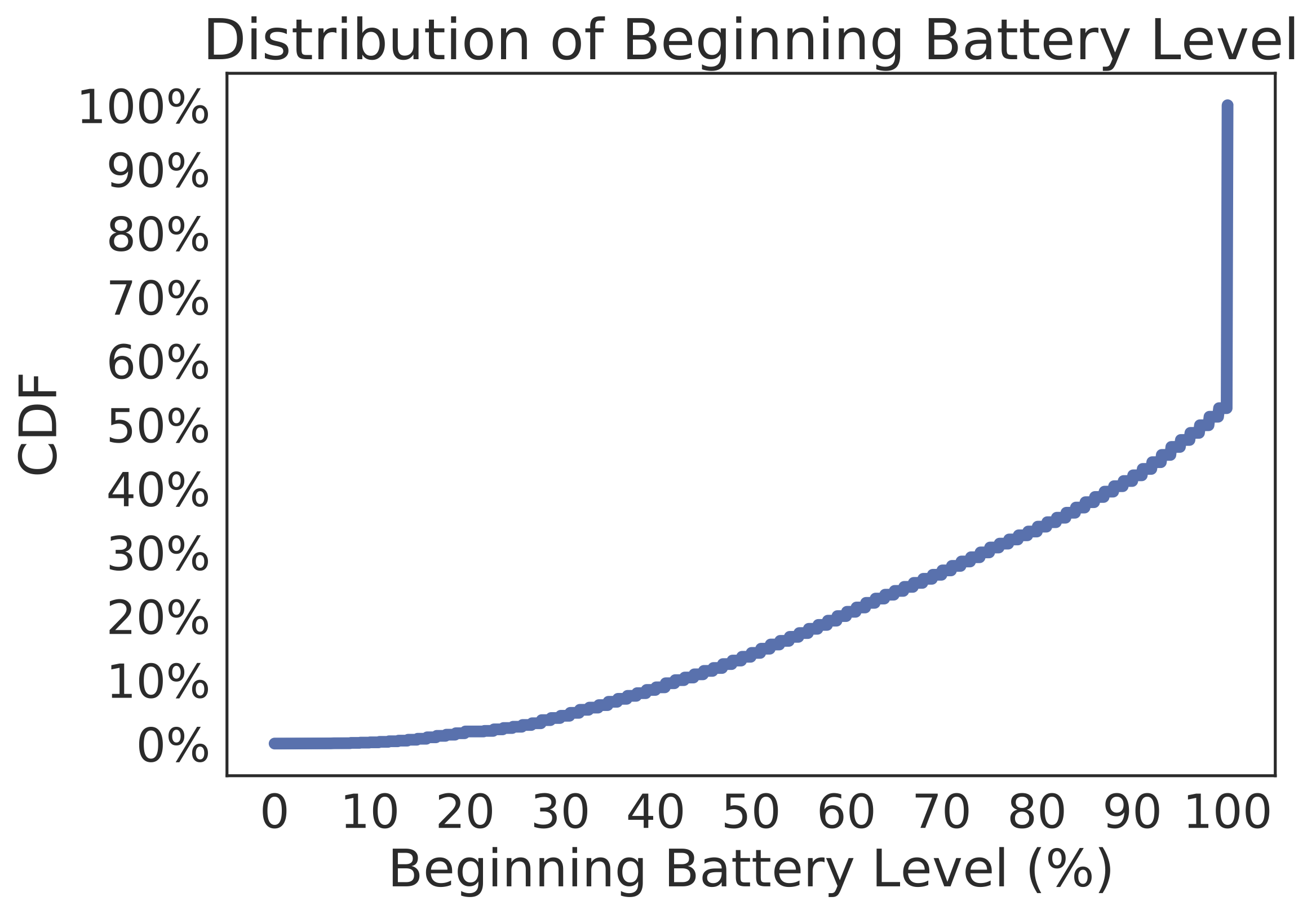}}
		\subfigure[Ending Battery Level\label{fig:cdf-battery-ending}]
		{\includegraphics[width=0.32\textwidth]{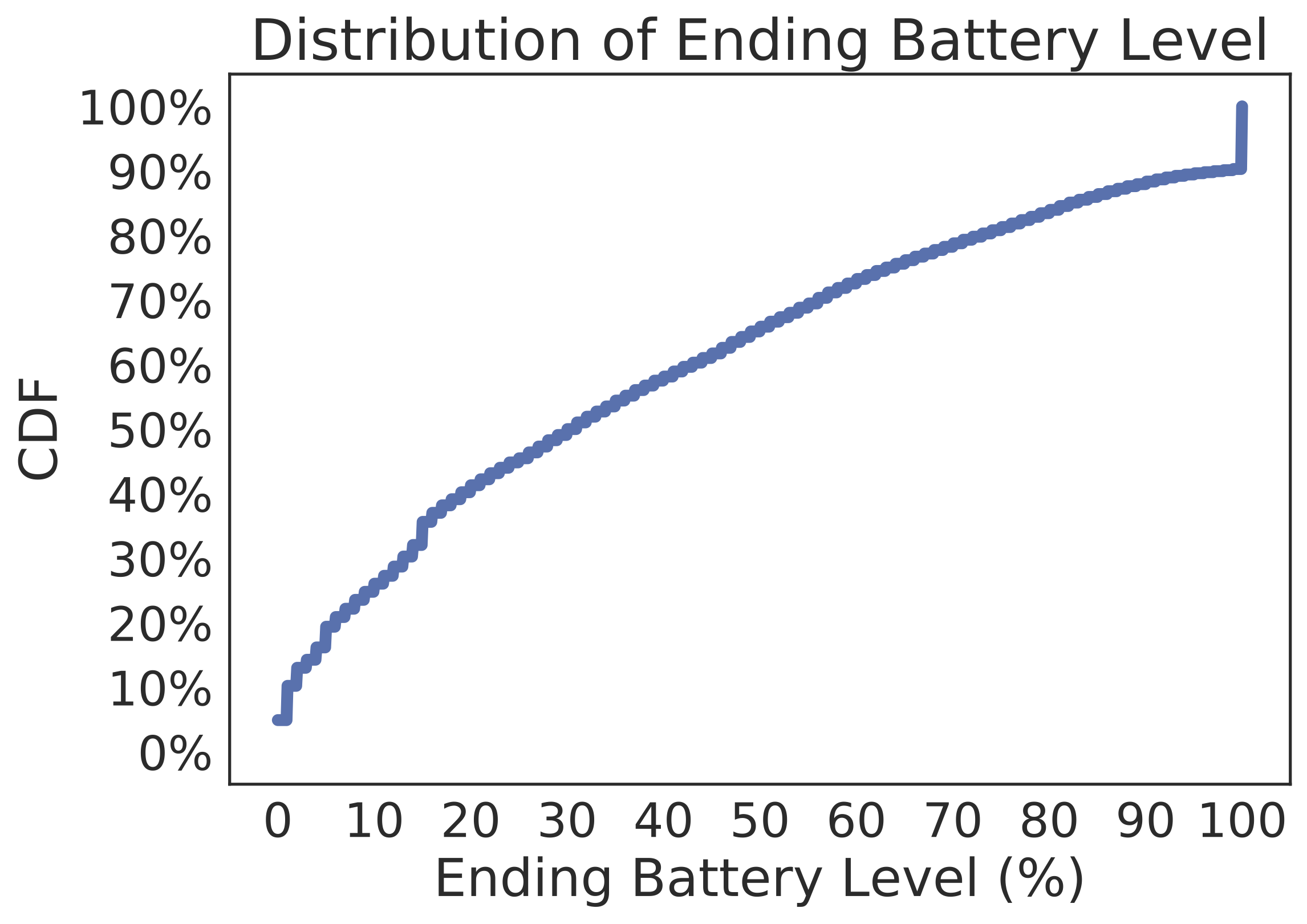}}
		\subfigure[Battery Consumption\label{fig:cdf-battery-usage}]
		{\includegraphics[width=0.32\textwidth]{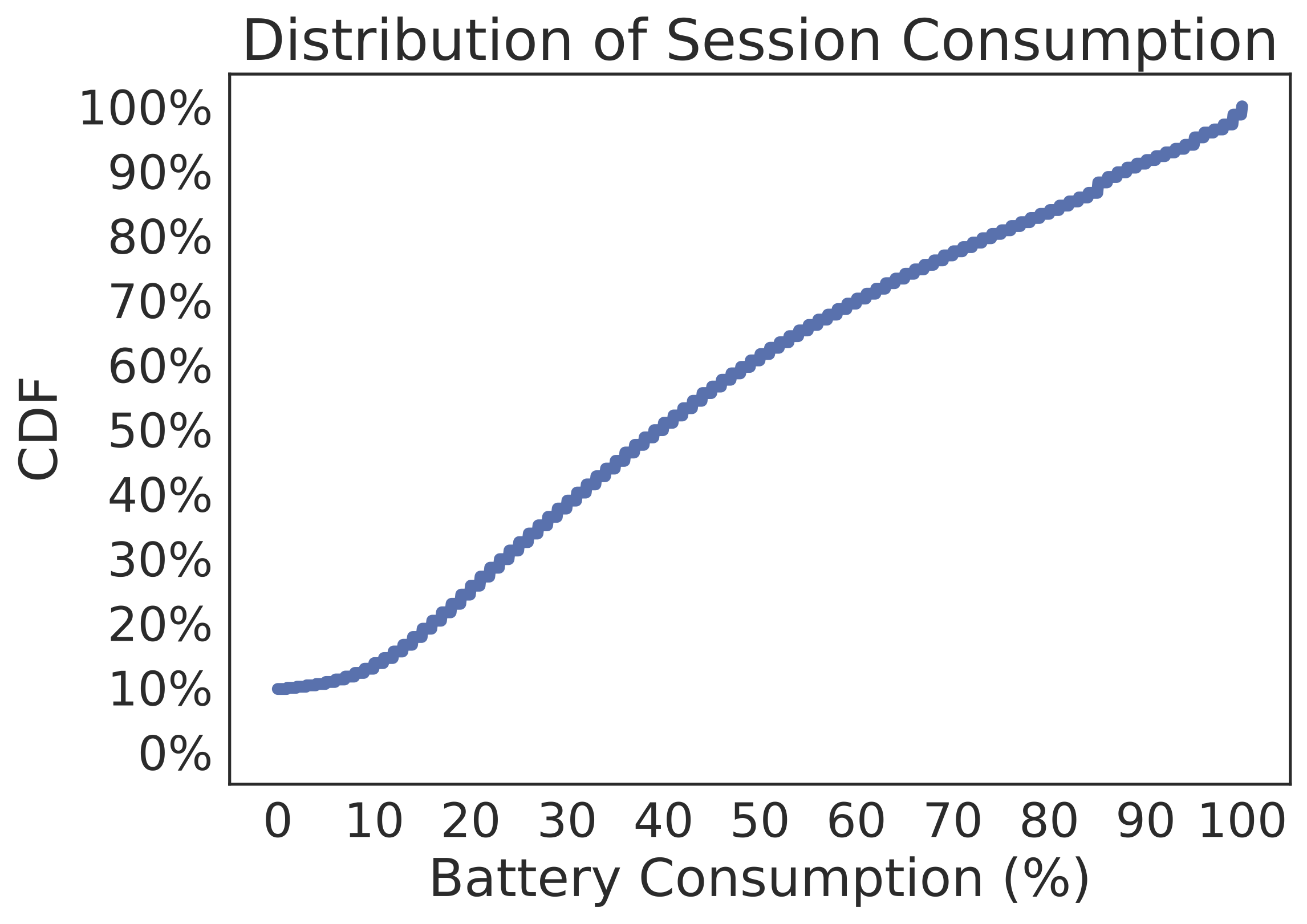}}
		\caption[7.5pt]{Cumulative distribution function (CDF) of begining battery level, ending battery level, and battery consumption.}\label{fig:cdf-battery}
	\end{center}
\end{figure}

Then, we investigate the distributions of battery usage within each session. The three figures in Figure~\ref{fig:cdf-battery} illustrate the distributions of the beginning battery level, the ending battery level, and the battery consumption, respectively. From Figure~\ref{fig:cdf-battery-beginning}, we can see that almost half of the sessions begin when the battery is totally full. This observation indicates that most users are used to completely charging their smartphones before they begin to use. For the rest of the sessions, the distribution of beginning battery level is close to uniform, indicating that users do not have any preferences on the time to begin their usage. 

The distribution of ending battery level is shown in Figure~\ref{fig:cdf-battery-ending}. This curve is pretty close to uniform distribution, too. Hence, we can know that users usually end their usage in different battery levels with similar possibilities. We then break down the two ends of the curve. As for the right end, we can see that about 10\% of the sessions end when the battery is completely full. We infer that these users prefer to keep their batteries fully charged at anytime, and they charge their smartphones whenever they can access available chargers, no matter the battery is full or not. On the other hand, 41\% of the sessions end when the battery level is lower than 20\%, and 5\% of the sessions ``die'', indicating that the battery totally runs out. 

Finally, we focus on the distribution of battery consumption demonstrated in Figure~\ref{fig:cdf-battery-usage}. Here, the battery consumption is defined as the difference between beginning battery level and ending battery level. Shown by the left bottom part of the figure, 10\% of the sessions do not consume any power. This part of sessions is almost the same group with the  sessions that end at the 100\% full status. For the other 90\% of sessions, the distribution of battery consumption is still close to a uniform distribution. In conclusion, the battery consumption of sessions has a large variance.

From the preceding analysis, almost all the observed distributions are consistent with our common-sense knowledge and can be interpreted, which demonstrates the validity of our data set.
\section{Experiment Setup}

Based on the definitions introduced in the last section, we then conduct a systematic experiment to demonstrate the predictive power of the features derived from our data set. In this section, we illustrate the workflow of the experiment.

\subsection{Workflow Overview}

\begin{figure}
	\centering
	\begin{center}
		\includegraphics[width=0.9\textwidth]{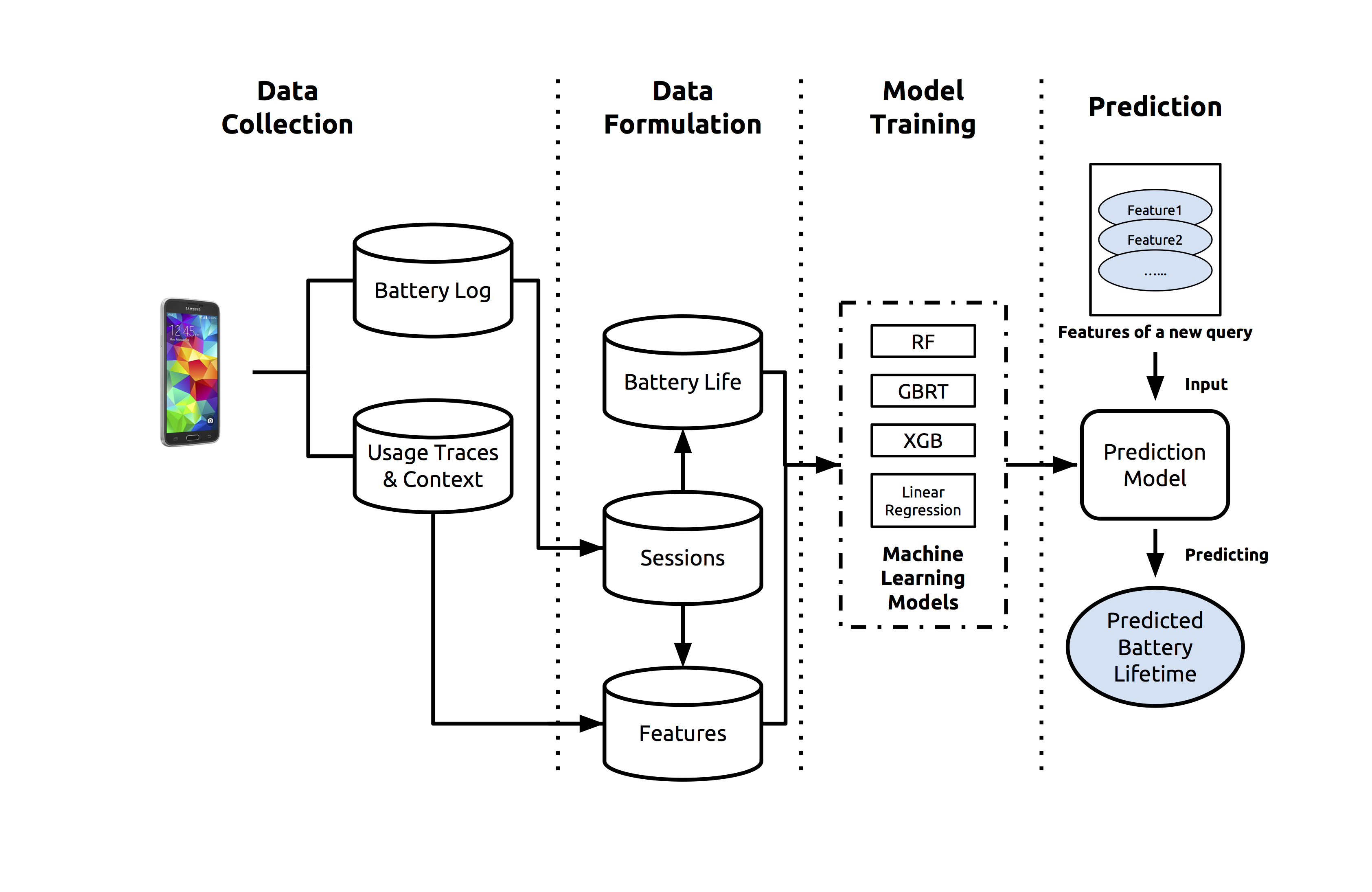}
		\caption[7.5pt]{An overview of the prediction model.}
		\label{fig:overview}
	\end{center}
\end{figure}

The general workflow of our experiment is illustrated in Figure~\ref{fig:overview}. In general, the battery life prediction model is generated in an offline fashion. Therefore, the model is trained based on a fixed data set in advance, and responds to new query instantly. The model consists of four major phases:

\begin{itemize}
    \item \textbf{Data Collection.} In the first phase, the model takes the usage data, system context, and the battery log from the Sherlock data set as input. The usage data includes system contexts, users' usage behavior, system events, etc. The battery log is the record of battery changes over time. 
    \item \textbf{Data Formulation.} We then process the raw data from the Sherlock data set to formulated features. First, we extract sessions from the battery log. Then we randomly sample time points in each session to simulate the user's query. Finally, by combining the session, the sampled time point, and the usage data, we could generate features of every single simulated query, and build a vector in a multi-dimensional space. The feature generation relies on only the information that occurs before the simulated query, so that building our model never ``looks into the future''.
    \item \textbf{Model Training.} The model then uses a group of regression models to derive the correlations between the features with the battery life. In this work, we synthesize the linear regression and three tree-based machine learning models. These models can capture both linear and non-linear correlations between the features and the battery life. This phase eventually generate a derived prediction model.
    \item \textbf{Prediction.} Finally, for a given new query, the model extracts the features, and then the derived model  predicts the remaining battery life.
\end{itemize}

\subsection{Evaluation Metrics}

In our experiment, we adopt three evaluation metrics: the Root Mean Square Error (RMSE), the Kendall's Tau, and the Concordance Index (C-idx). The measurement of these metrics and the reasons why we choose them are presented as follows.

\subsubsection{Root Mean Square Error (RMSE)}

The most critical issue in our experiment is the evaluation metric of the predicted outcome. The most straightforward way is calculating the absolute gap between the actual battery life and the predicted battery life. Then, by using standard statistical methods such as \textbf{\textit{Root Mean Square Error}} (RMSE), we can measure the prediction error. In practice, RMSE is widely used for regression problems, and it can provide an intuitive illustration of the absolute prediction error.

Although RMSE is commonly used in regression problems, it inherently has some non-negligible limitations. For example, RMSE is quite sensitive to outliers, and it will be significantly influenced by the  scale of data~\cite{armstrong2001evaluating}. More seriously, when applied to forecast problems, RMSE has the disadvantage that it puts a heavier penalty on positive errors than on negative errors. In other words, it is not symmetric for error evaluation~\cite{hyndman2006another}. Therefore, RMSE is not always reliable enough if we take RMSE as the only metric. 

\subsubsection{Kendall's Tau}

To address the limitations mentioned above, we take into account the ranking-based methods to measure how much the ranking of the predicted life is consistent with the ranking of the actual life. Statistically, such consistency can be measured by a widely used metric namely \textbf{\textit{Kendall's Tau}}~\cite{manning2008introduction}. Kendall's Tau measures the difference between the number of normalized concordant pairs and the number of normalized discordant pairs in two ranking lists:

\[\tau=\frac{\Sigma_{i,j}(\mathds{I}[(x_i-x_j)\cdot(y_i-y_j)>0]-\mathds{I}[(x_i-x_j)\cdot(y_i-y_j)<0])}{\Sigma_{i,j}(\mathds{I}[(x_i-x_j)\cdot(y_i-y_j)>0]+\mathds{I}[(x_i-x_j)\cdot(y_i-y_j)<0])}\]

where $i,j$ are two sessions, $x_i, y_i$ are the orders of the two sessions into two ranking lists. $\mathds{I}[\cdot]$ is an indicator, which equals to 1 if the inside expression is true and 0 otherwise. $\tau$ shall be 1 if the two ranking lists perfectly match, -1 if they totally disagree with each other, and 0 if they are entirely independent. Since every session is considered equally, Kendall's Tau will not be very sensitive to extreme outliers, so it is a meaningful supplement to the RMSE.

\subsubsection{Concordance Index (C-idx)}

RMSE and Kendall's Tau can measure the predicting accuracy from different perspectives and they can complement each other. However, the basic precondition that these two metrics can be applied is both the actual and predicted battery life can be explicitly observed. Such a precondition is too strong to be always true. As we discussed in Section~\ref{sec:low-battery-event-and-battery-lifetime}, a low-battery event may not happen in all sessions. For sessions that end before reaching the threshold $L$, we can hardly know their actual battery life. In this case, either RMSE or Kendall's Tau does not work. From Figure~\ref{fig:cdf-battery-ending}, we can learn that most sessions' actual battery life cannot be known if $L$ is close to 0. In other words, some essential data is missing and we need to handle such a problem to ensure the completeness and correctness of our prediction task. 

We then try to partially address the data-missing problem by employing a metric called \textbf{\textit{Concordance Index}} (C-idx)~\cite{harrell1982evaluating}. C-idx is a widely used metric in \textbf{survival analysis}. The goal of survival analysis is to analyze the expected duration of time until one or more events happen\footnote{\url{https://en.wikipedia.org/wiki/Survival_analysis}}. For example, the expected survival time for a group of lung-cancer patients. In survival analysis, a subject is called \textbf{censored} if it does not have any event during the observation time. In the lung-cancer patients scenario, a patient is censored if the event (i.e., death) does not happen before they quits the experiment. C-idx can then deal with the data sets that contain censored data points. For each pair of subjects, C-idx estimates the probability that the predicted outcome is consistent with the ground truth. For example, if a patient A passes away in 3 months, and patient B is censored at 5 months, then we can be confident that B survives longer than A even if we do not know B's actual survival time. Hence, the probability can be explicitly obtained in this case. In another case, if A passes away in 5 months and B is censored at 3 months, then we can not clearly know which patient survives longer, and the probability is 0.5. The overall average of all pairs' probability is the final result of C-idx, whose value is 1 if the two lists are perfectly concordant and 0 when they are entirely anti-concordant. Particularly, C-idx is linearly equivalent to Kendall's Tau when all data points are not censored. Therefore, C-idx can be considered as a generalization of Kendall's Tau and can alleviate the data-missing problem. Inherently, our battery life prediction task is quite similar to a survival analysis problem, so we can employ C-idx as a promising metric for our problem. 

In conclusion, RMSE, Kendall's Tau, and C-idx could measure the prediction's accuracy from different perspectives. RMSE could measure the absolute error of the prediction so it can provide an intuitive insight of the accuracy. Kendall's Tau is rank-based and widely adopted, which could show the prediction accuracy on uncensored data. C-idx is used to demonstrate the model's predictive ability on both censored and uncensored data points. Therefore, when evaluating the models, we will take all these there metrics. To be specific, we apply RMSE and Kendall's Tau to observed sessions, and apply C-idx to all sessions.

\subsection{Data Cleaning and Feature Selection}

In this section, we describe how we parse the raw data and transform the raw data to the features that can be fed to the prediction model.

\subsubsection{Query Simulation}

We generate our experiment data from the 37,088 sessions mentioned in~\ref{sec:session-generation}. First of all, sessions whose beginning battery level is lower than 30\% are discarded. This is because these sessions are already in a low battery state when they begin, so predicting their battery life is not meaningful. Such a step filters out about 4\% sessions, and keeps 35,590 sessions left. We fit the threshold $L$ to 20\%, under which 13,825 sessions are observed ones and the rest 21,765 sessions are unobserved ones. 

For every single session, we assume that a user can issue a query of battery life at any time. Therefore, we randomly sample a time point $t_{query}$ that satisfies $t_{query}\in[t_{start}+2min, t_{event}-2min]$ for observed sessions and $t_{query}\in[t_{start}+2min, t_{end}-2min]$ for unobserved sessions. To make sure that there already exist some usage data before the prediction query, we add a 2-minute boarder at the beginning so that the sample time-points are too close to the beginning. Similarly, we also add a 2-minute boarder at the end to make sure that the sample time points are not too close to the end.

We then sample 1/6 sessions from both observed sessions and unobserved sessions as our testing set. Other sessions are used for training the model. The summary of statistics of the data is listed in Table~\ref{tab:data-stats}.

\begin{table}[htbp]
\centering
\caption{Summary statistics of the data.}
\label{tab:data-stats}
\begin{tabular}{|l|rr|r|}
\hline
 & \textbf{Observed} & \textbf{Unobserved} & \textbf{Total} \\
\hline
\textbf{Training Data} & 11,520 & 18,137 & 29,657 \\
\textbf{Testing Data} & 2,305 & 3,628 & 5,933 \\
\hline
\textbf{Total} & 13,825 & 21,765 & 35,590 \\
\hline
\end{tabular}
\end{table}

\subsubsection{Feature Extraction}

Since the Sherlock data set has a great number of dimensions, it is unnecessary and unrealistic for us to use all of them in the experiment. For simplicity, we use only the following four parts of the Sherlock data set to extract features: app usage data, sensor data, screen status, and broadcast record. We select these parts due to the following reasons. App usage data could reflect users' usage behavior, while sensor data could illustrate the context of users' usage, so it is natural to consider these two parts. As for screen status and broadcast, they are the most basic system events which can be easily recorded, and they are user perceivable. In addition, since other system events are too rare or have other quality problems, we choose only these two parts as a representation of the system events.

The description of each part is listed as follows:

\begin{itemize}
    \item \textbf{App usage data.} The app usage data are collected through periodical sampling. The agent collects apps' running information every 5 seconds. The raw data of app usage contains various fields. In our model, we use only the ``status'' fields, that is, we consider only the app's running status at a specific time point, such as running in the foreground, idle, and so on.
    \item \textbf{Sensor data.} According to the categorization of the Sherlock data set, sensors on the Samsung Galaxy S5 smartphones are categorized into 4 groups, named from T1 to T4. Due to the consideration of data completeness, we select 9 fields from group T1 and 150 fields from group T2 in our model. T1 and T2 are periodically sampled from the system  once a minute and once every 15 seconds, respectively.
    \item \textbf{Screen status and broadcast record.} Unlike the previous 2 parts of data, screen status and broadcast record are collected through the event-base strategy. For screen status, each time the user turns screen on or off, the agent will record this event with its timestamp. Similarly, whenever the system sends a broadcast, the agent can record the timestamp and the type of the broadcast. 
\end{itemize}

Based on the data above, we finally extract 21 groups of features, which are described in Table~\ref{tab:features}. The detailed explanations on these features will be introduced in the following section. To ensure that all features have similar scale, we add the standard normalization to each feature, that is, adjusting their mean to 0 and standard deviation to 1 through a linear transformation. We use the median of the valid values of every single data field to fill the missing values in this field.

\subsection{Baseline: Ranking by Remaining Battery Volume}

An intuitive and straightforward way to estimate the battery life is based on the current battery level. If the battery has more power, it should last longer in most cases. To demonstrate the predictive ability of this fashion, we first train the regression models based on only the battery level at $t_{query}$ (F1), and evaluate the models' performance on testing sets. We explore a group of standard, state-of-the-art machine-learning models for this problem, including the Linear Regression, the Random Forest Regression (RF)~\cite{breiman2001random}, the Gradient Boosted Regression Tree (GBRT)~\cite{friedman2002stochastic}, and the Extreme Gradient Boosting (XGBoost)~\cite{friedman2001greedy}. These models can capture both linear and non-linear correlations\footnote{We use the XGBoost package~\cite{chen2016xgboost} for XGBoost, and scikit-learn package~\cite{pedregosa2011scikit} for Linear Regression, RF, and GBRT. Both packages are implemented in Python. We use default parameters when training all the models. For example, \texttt{n\_estimators=100} and \texttt{learning\_rate=0.1} for GBRT and XGBoost, while \texttt{max\_features=''auto''} for RF}. We have also considered two additional models: SVM Regression and Decision Tree Regression. Their performance is inferior to the aforementioned models. For simplicity, we will focus on the four models in this paper.

A smartphone is actually a  computer system that contains numerous software and hardware components. The battery consuming rate is not likely to be affected by only linear combinations of these factors. From this point of view, we think that it is useful to capture inner correlations among system features,  and non-linear correlations between system features and battery consuming rate. It is the reason why we adopt these tree-based machine learning models, and \textbf{we make the hypothesis that tree-based models can outperform the linear regression model}. Generally, the three tree-based models are state-of-the-art models that are all widely used. We will compare their performance when evaluating the prediction results.

The results are presented in the first three rows of Table~\ref{tab:result-current}. These simple models yield a RMSE of 140.4 and a $\tau$ of 0.7442 on the 2,305 observed sessions. Results generated from all the four models are quite similar, which is reasonable because there is only one variable. The result of RMSE is not satisfying, implying that the estimated error of this simple model is about 140 minutes. Even though, it is surprising to see that the result of ranking does not seem that bad. The $\tau$ value is about 0.7 and means that there is a strong positive correlation between current battery level and battery life. Although the absolute error of this method is a bit high, the order of sessions according to the battery life is promising in most cases. When calculating the C-idx on all 5,933 testing sessions, the highest result is 0.9260. Such a result is quite similar to Kendall's Tau. 

The inconsistency between the RMSE and the ranking-based metrics can be informative: the current battery level is quite crucial to the battery life,  but its actual predictive ability is not satisfying at all. We are motivated by this finding to figure out whether and how much the other system status and usage behavior can further improve the predicting accuracy. 
\section{Experiment Results}

\begin{figure}
	\centering
	\begin{center}
		\includegraphics[width=0.8\textwidth]{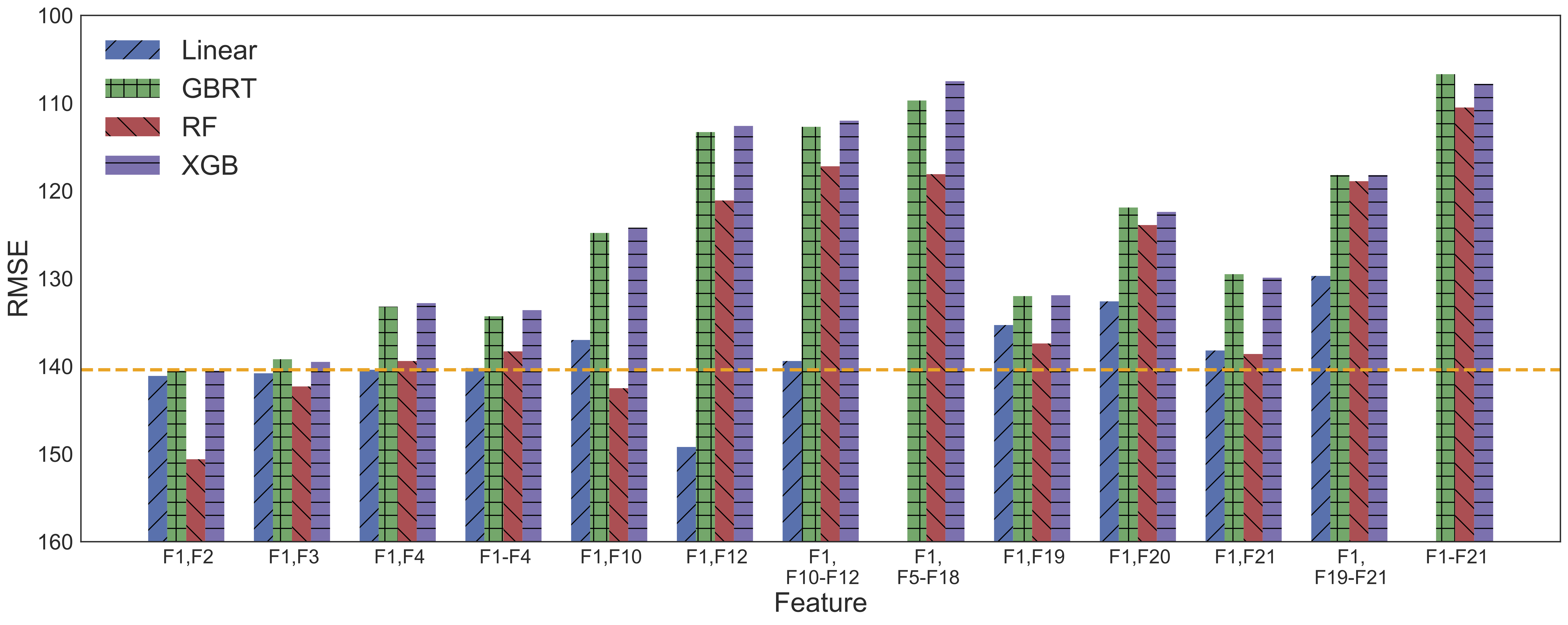}
		\caption[7.5pt]{Select results of RMSE. The orange dashed line represents the baseline.}
		\label{fig:result-rmse}
	\end{center}
\end{figure}

\begin{figure}
	\centering
	\begin{center}
		\includegraphics[width=0.8\textwidth]{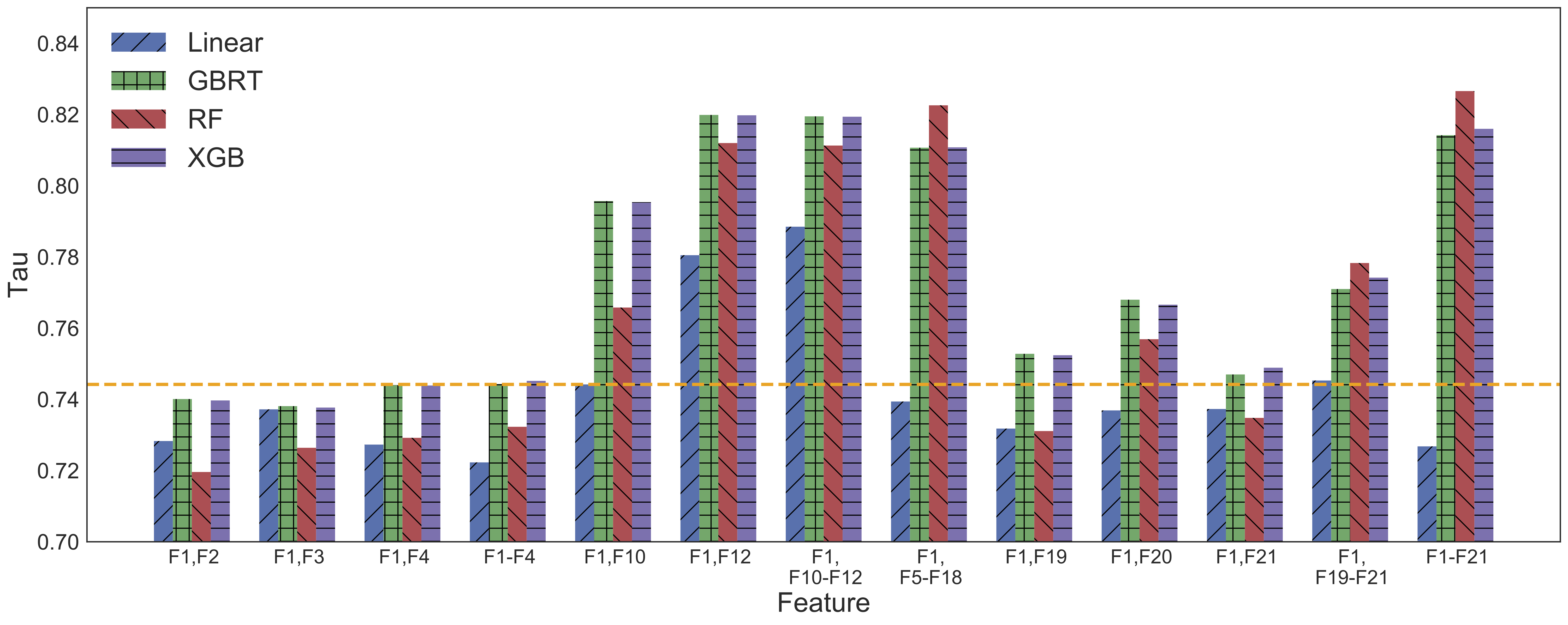}
		\caption[7.5pt]{Select results of Tau. The orange dashed line represents the baseline.}
		\label{fig:result-tau}
	\end{center}
\end{figure}

In this section, we present the predictive power of each group of feature, and whether the prediction accuracy could be enhanced through the combination of features. During this section, we will also give some discussions about the features' performance. The performance of some selected feature groups is demonstrated through barplots (Figure~\ref{fig:result-rmse}, Figure~\ref{fig:result-tau}, and Figure~\ref{fig:result-cidx}) and the complete results are organized in the tables.

\newcolumntype{L}[1]{>{\raggedright\let\newline\\\arraybackslash\hspace{0pt}}m{#1}}
\newcolumntype{C}[1]{>{\centering\let\newline\\\arraybackslash\hspace{0pt}}m{#1}}
\newcolumntype{R}[1]{>{\raggedleft\let\newline\\\arraybackslash\hspace{0pt}}m{#1}}

\small
\begin{table}[htbp]
\centering
\caption{Features of the Regression Models. Features with a star (*) are represented in \underline{one-hot} format: if the features has $k$ possible discrete values, then this feature is represented by $k$ fields. The corresponding field equals to 1 while other fields equal to 0. We use the median of the valid values of each data field to fill the missing values in this field.}
\label{tab:features}
\begin{tabular}{|l|l|R{1.5cm}|L{9cm}|}
\hline
\textbf{Group} & \textbf{Name} & \textbf{\# of Features} & \textbf{Description} \\
\hline
F1 & current\_battery & 1 & The battery level of the query time. \\
\hline
F2* & current\_hour & 24 & The hour of the query time. (0-23) \\
\hline
F3* & current\_weekday & 7 & The weekday of the query time. (0-6) \\
\hline
F4 & sensor\_T2\_last & 150 & The average value of the 150 T2 sensors of the last available entry. \\
\hline
F5 & start\_battery & 1 & The battery level of the session's beginning time. \\
\hline
F6* & start\_hour & 24 & The hour of the session's beginning time. (0-23) \\
\hline
F7* & start\_weekday & 7 & The weekday of the session's beginning time. (0-6) \\
\hline
F8 & age & 1 & The duration from the session's beginning to the query time. \\
\hline
F9 & consumption & 1 & The consumed battery from the session's beginning to the query time. \\
\hline
F10 & history\_rate & 1 & The average battery consuming rate in the past of the session. (F10=F9/F8) \\
\hline
F11 & naive\_surv & 1 & The na\"ive prediction of the battery life based on F1 and F10. (F11=F1/F10) \\
\hline
F12 & past\_rate & 11 & The average battery consuming rate when consume the most recent 1\%-10\% battery. The time period that the battery stays in the current battery level. \\
\hline
F13 & sensor\_T1 & 45 & The average value of the nine T1 sensors in the past 1/5/10/30/60 minutes. \\
\hline
F14 & sensor\_T2\_5min & 150 & The average value of the 150 T2 sensors in the past five minutes. \\
\hline
F15 & app\_occurrence & 400 & The binary indicators of the top 50 apps' occurrence in the past 5/10/30/60 minutes. 400 features=4 threshold * 50 apps * 2 states (foreground \& background). \\
\hline
F16 & app\_usage & 100 & The percentage of the top 50 apps' usage time in the past of the session. 100 features=50 apps * 2 states (foreground \& background). \\
\hline
F17 & screen & 2 & \# of screen on actions \& the percentage of screen on time. \\
\hline
F18 & broadcast & 86 & \# of the received broadcasts of all 86 broadcasts. \\
\hline
F19* & user\_index & 51 & User's unique index (0-50) \\
\hline
F20 & session\_history & 8 & The statistics of battery consuming rate of this user's historical sessions. There are four types of session filtering rules when extracting this feature: consider sessions begin at 1) anytime, 2) in the same hour, 3) in the same weekday, and 4) in the same hour and weekday. Mean and median are both considered for each rule, so there are 8 features in total. \\
\hline
F21 & screen\_history & 8 & The statistics of screen on actions of this user's historical sessions (Similar to F20). \\
\hline
\end{tabular}
\end{table}
\normalsize

\subsection{Query-time Feature}

Beyond the battery level, the contexts of the query can also be used to predict the battery life. We extract four groups of features that could describe the contexts: current battery level (F1), current hour (F2), current weekday (F3), and T2 sensors (F4). We call these features \textbf{``query-time feature''} because they can only illustrate the information when the query occurs. F1 is what we used to build the baseline. F2 and F3 are the coarse-grained description of the current time. F4 can demonstrate the system status of the device. We first try to build regression models based on each single group of features, by which we finally got a set of terrible results. The best $\tau$ and RMSE are only 0.2791 and 190.2, respectively, which are much worse than the baseline. This conveys to us that the context itself cannot make accurate prediction alone. Due to this reason, we always bind other groups of sessions with F1 to train the model, to see if the baseline can be improved.

We separately combine F2-F4 with F1 to train the model, and finally put all these four groups together to check the performance. All the results are shown in Table~\ref{tab:result-current}. Comparing to the baseline, adding query-time features almost has no improvement at all, even if putting all the features together. RMSE is improved from 140.4 to 132.8, but the other two metrics' improvements are too tiny. In conclusion, the context can slightly reduce the absolute error of the prediction, but it has no effect on the ranking of the sessions. In other words, the battery level itself is quite dominant if we only consider the information at the query time.

\begin{table}[htbp]
\centering
\caption{Performance of query-time features. Best performances under each metric are highlighted.}
\label{tab:result-current}
\begin{tabular}{|l|l|r|r|r|r|}
\hline
\multirow{2}{*}{\textbf{Feature}} & \multirow{2}{*}{\textbf{Metrics}} & \multicolumn{4}{c|}{\textbf{Model}} \\
\cline{3-6}
 & & \multicolumn{1}{c|}{Linear} & \multicolumn{1}{c|}{GBRT} & \multicolumn{1}{c|}{RF} & \multicolumn{1}{c|}{XGB} \\
\hline
\multirow{3}{*}{F1} & RMSE & 141.2 & 140.5 & 140.7 & 140.4 \\
 & Tau & 0.7442 & 0.7435 & 0.7433 & 0.7436 \\
 & C-idx & 0.9260 & 0.9257 & 0.9255 & 0.9257 \\
\hline
\multirow{3}{*}{F1, F2} & RMSE & 141.1 & 140.5 & 150.6 & 140.5 \\
 & Tau & 0.7283 & 0.7401 & 0.7196 & 0.7397 \\
 & C-idx & 0.9241 & \textbf{0.9274} & 0.9194 & 0.9273 \\
\hline
\multirow{3}{*}{F1, F3} & RMSE & 140.8 & 139.2 & 142.3 & 139.5 \\
 & Tau & 0.7372 & 0.7381 & 0.7264 & 0.7377 \\
 & C-idx & 0.9260 & 0.9261 & 0.9214 & 0.9260 \\
\hline
\multirow{3}{*}{F1, F4} & RMSE & 140.4 & 133.2 & 139.4 & \textbf{132.8} \\
 & Tau & 0.7273 & 0.7440 & 0.7292 & 0.7439 \\
 & C-idx & 0.9235 & 0.9291 & 0.9223 & 0.9292 \\
\hline
\multirow{3}{*}{F1-F4} & RMSE & 140.2 & 134.3 & 138.3 & 133.6 \\
 & Tau & 0.7223 & 0.7443 & 0.7323 & \textbf{0.7452} \\
 & C-idx & 0.9226 & 0.9296 & 0.9239 & 0.9298 \\
\hline
\end{tabular}
\end{table}

\subsection{Session Feature}

The unsuccessful attempt of query-time features tells us the predicting power of query time's context is poor. Even though, this result is not out of our expectation, and the phenomenon is reasonable and easy to understand. The query-time features could only reflect the device's status at a single time point, but it cannot tell the usage behavior and the system status in the future at all, which is more determinant to the battery life. Therefore, it should be very helpful if we could predict the user's usage and the system status in the future. However, this task is complicated and beyond the scope of this study. We need to find other simple ways to estimate it. We have the hypothesis that users' usage will usually be stable. Their usage behavior will not significantly change within a single session. In this case, the usage behavior and system status before the query time can somehow reflect the usage pattern in the future, so it can help predicting the battery life. 

Motivated by the discussion above, we are eager to see how well the model can perform based on the usage history within the session. We call this part of features \textbf{``session feature''}. We make a comprehensive extraction of session features, and finally we get 14 groups (F5-F18 in Table~\ref{tab:features}). These features include the context when the session begins (F5-F7), the duration and power consuming in the past (F8-F12), app usage history and aggregated value of sensors (F13-F16), and system events occurred within the session (F17, F18). Similarly, we first separately combine each group of sessions with F1 to see whether they can improve the performance alone. The results are presented in the upper part of Table~\ref{tab:result-session}. 

\begin{figure}
	\centering
	\begin{center}
		\includegraphics[width=0.8\textwidth]{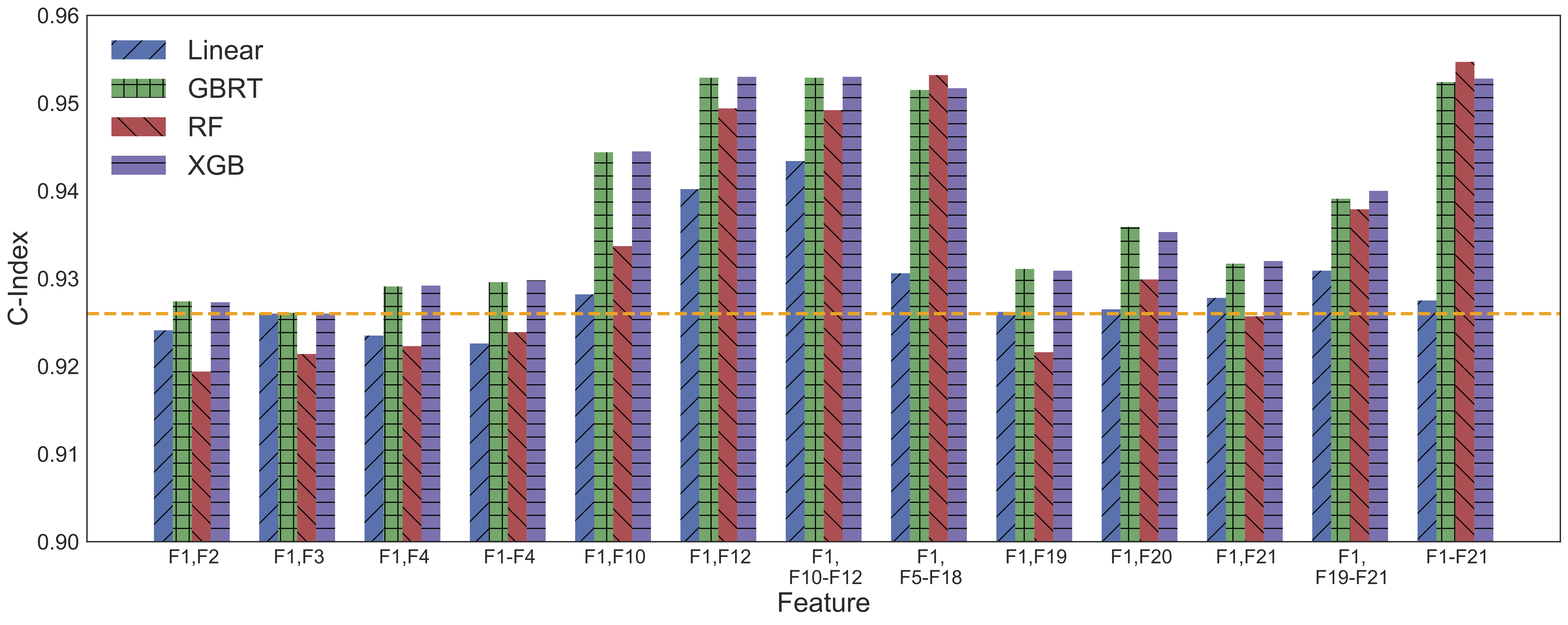}
		\caption[7.5pt]{Select results of C-idx. The orange dashed line represents the baseline.}
		\label{fig:result-cidx}
	\end{center}
\end{figure}

\begin{table}[htbp]
\centering
\caption{Performance of session features. Significant improvements (RMSE $\leq125$, $\tau\geq 0.75$, C-idx $\geq 0.94$) are highlighted.}
\label{tab:result-session}
\footnotesize
\begin{tabular}{|l|l|r|r|r|r|l|l|r|r|r|r|}
\hline
\multirow{2}{*}{\textbf{Feature}} & \multirow{2}{*}{\textbf{Metrics}} & \multicolumn{4}{c|}{\textbf{Model}} & \multirow{2}{*}{\textbf{Feature}} & \multirow{2}{*}{\textbf{Metrics}} & \multicolumn{4}{c|}{\textbf{Model}} \\
\cline{3-6}\cline{9-12}
 & & \multicolumn{1}{c|}{Linear} & \multicolumn{1}{c|}{GBRT} & \multicolumn{1}{c|}{RF} & \multicolumn{1}{c|}{XGB} & & & \multicolumn{1}{c|}{Linear} & \multicolumn{1}{c|}{GBRT} & \multicolumn{1}{c|}{RF} & \multicolumn{1}{c|}{XGB} \\

\hline
\multirow{3}{*}{F1, F5} & RMSE & 141.2 & 139.9 & 144.4 & 139.8 & \multirow{3}{*}{F1, F12} & RMSE & 149.2 & \textbf{113.3} & \textbf{121.1} & \textbf{112.6} \\
& Tau & 0.7394 & 0.7400 & 0.7053 & 0.7400 & & Tau & \textbf{0.7805} & \textbf{0.8199} & \textbf{0.8120} & \textbf{0.8198} \\
& C-idx & 0.9262 & 0.9254 & 0.9136 & 0.9255 & & C-idx & \textbf{0.9402} & \textbf{0.9529} & \textbf{0.9494} & \textbf{0.9530} \\
\hline
\multirow{3}{*}{F1, F6} & RMSE & 140.0 & 136.6 & 144.9 & 136.4 & \multirow{3}{*}{F1, F13} & RMSE & 140.5 & 134.5 & 144.3 & 134.6 \\
& Tau & 0.7278 & 0.7384 & 0.7143 & 0.7392 & & Tau & 0.7343 & 0.7413 & 0.7224 & 0.7443 \\
& C-idx & 0.9238 & 0.9268 & 0.9152 & 0.9271 & & C-idx & 0.9259 & 0.9284 & 0.9197 & 0.9290 \\
\hline
\multirow{3}{*}{F1, F7} & RMSE & 140.6 & 139.6 & 143.5 & 139.6 & \multirow{3}{*}{F1, F14} & RMSE & 139.5 & 131.7 & 137.9 & 132.1 \\
& Tau & 0.7370 & 0.7380 & 0.7269 & 0.7384 & & Tau & 0.7268 & 0.7414 & 0.7378 & 0.7425 \\
& C-idx & 0.9259 & 0.9260 & 0.9215 & 0.9261 & & C-idx & 0.9237 & 0.9287 & 0.9242 & 0.9290 \\
\hline
\multirow{3}{*}{F1, F8} & RMSE & 136.5 & 129.4 & 144.2 & 129.6 & \multirow{3}{*}{F1, F15} & RMSE & HUGE & 131.8 & 137.4 & 131.5 \\
& Tau & 0.7434 & \textbf{0.7613} & 0.7219 & \textbf{0.7624} & & Tau & 0.7249 & 0.7453 & 0.7452 & 0.7471 \\
& C-idx & 0.9297 & 0.9348 & 0.9207 & 0.9351 & & C-idx & 0.9234 & 0.9297 & 0.9279 & 0.9301 \\
\hline
\multirow{3}{*}{F1, F9} & RMSE & 141.2 & 139.8 & 143.6 & 139.8 & \multirow{3}{*}{F1, F16} & RMSE & 136.2 & 130.8 & 138.7 & 129.4 \\
& Tau & 0.7394 & 0.7398 & 0.7099 & 0.7398 & & Tau & 0.7283 & 0.7449 & \textbf{0.7522} & 0.7458 \\
& C-idx & 0.9262 & 0.9255 & 0.9150 & 0.9255 & & C-idx & 0.9256 & 0.9304 & 0.9307 & 0.9307 \\
\hline
\multirow{3}{*}{F1, F10} & RMSE & 137.0 & \textbf{124.8} & 142.5 & \textbf{124.2} & \multirow{3}{*}{F1 ,F17} & RMSE & 139.5 & 135.4 & 144.2 & 134.9 \\
& Tau & 0.7442 & \textbf{0.7957} & \textbf{0.7658} & \textbf{0.7954} & & Tau & 0.7404 & 0.7496 & 0.7207 & \textbf{0.7506} \\
& C-idx & 0.9282 & \textbf{0.9444} & 0.9337 & \textbf{0.9445} & & C-idx & 0.9277 & 0.9306 & 0.9198 & 0.9308 \\
\hline
\multirow{3}{*}{F1, F11} & RMSE & 137.3 & 125.5 & 141.6 & \textbf{124.9} & \multirow{3}{*}{F1, F18} & RMSE & 145.7 & 130.8 & 133.8 & 129.7 \\
& Tau & 0.7439 & \textbf{0.7952} & \textbf{0.7665} & \textbf{0.7948} & & Tau & 0.7406 & \textbf{0.7558} & 0.7497 & \textbf{0.7579} \\
& C-idx & 0.9280 & \textbf{0.9444} & 0.9335 & \textbf{0.9443} & & C-idx & 0.9283 & 0.9343 & 0.9310 & 0.9349 \\

\hline
\multicolumn{12}{|c|}{Putting together} \\
\hline
\multirow{3}{*}{F1, F10-F12} & RMSE & 139.4 & \textbf{112.7} & \textbf{117.2} & \textbf{112.0} & \multirow{3}{*}{F1, F5-F18} & RMSE & HUGE & \textbf{109.7} & \textbf{118.1} & \textbf{107.5} \\
& Tau & \textbf{0.7885} & \textbf{0.8195} & \textbf{0.8113} & \textbf{0.8194} & & Tau & 0.7394 & \textbf{0.8107} & \textbf{0.8226} & \textbf{0.8108} \\
& C-idx & \textbf{0.9434} & \textbf{0.9529} & 0.9492 & \textbf{0.9530} & & C-idx & 0.9306 & \textbf{0.9515} & \textbf{0.9532} & \textbf{0.9517} \\

\hline
\end{tabular}
\end{table}
\normalsize

Compared to query-time features, some session features do have the significant predictive power. From Table~\ref{tab:result-session}, we can see that GBRT and XGB on F10 and F11 could enhance the predicting performance. F10 is the average battery consuming rate in the past of the session. For example, if the session has lasted for 2 hours and it consumes 40\% of battery, then F10 is equal to $40\%/2\ hours=20$. F11 is the naive prediction based on F10, which under the assumption that the battery consuming rate will always stay at F10. As described in Table~\ref{tab:features}, we have F11=F1/F10. These two features are quite useful even though they are simple. The best RMSE has a reduction of 16 minutes, while the best $\tau$ can be improved by 0.05. RF works worse than the other two tree-based regression models. It cannot effectively reduce the RMSE, however, it still can slightly increase the $\tau$. In other words, RF cannot capture the absolute error well, but it could help to make the right order of sessions based on F10 and F11.

If we use the battery discharging information in a more detailed way, the results can be better. The most powerful group of session features is the past consuming rate (F12). The meaning of F12 is the average power consuming rate of the past 1\%-10\% of battery before the query time. For example, if it takes 50 minutes to consume the recent 5\% of battery, then the corresponding field is $50/10=10$ minutes. Except for the RMSE of the linear regression's result which is worse than the baseline, the performance of all models are significantly raised under all metrics. The best results of a single group so far are also generated by F12, which are RMSE\ $=112.6$, $\tau=0.8120$, and C-idx\ $=0.9530$. The more informative battery consumption history provides us with more insights about user's usage pattern, so it is more predictive than the coarse-grained ones. 

Except for F10, F11, and F12, other features are not informative enough. Tree-based methods can raise $\tau$ on age of the session (F8), app usage (F16), screen actions (F17), and broadcasts (F18), though the improvement is little. For other groups, there is no improvement at all. 

\textbf{Discussion.} According to the analysis, only the 3 groups of features related to battery discharging history have significantly predictive ability. Other features including context, usage history, and device status are almost not useful at all. This finding is interesting. An explanation is, no matter how users use the device and what the device status is, their influence on battery will finally be reflected through the discharging rate. The discharging pattern is the most fundamental and straightforward aspect.

On the other hand, if comparing different regression models, we can find that linear regression almost has no predictive power. Therefore, the correlation between the features and the outcome should be non-linear, which can only be captured by tree-based methods. Another serious problem of linear regression is over-fitting. When applying it to F15, the RMSE of the outputs is non-sense large ($>1e5$) because the model is seriously over-fitting on this group of features. This is because F15 contains a large group of sensors (400 different indicators), and the variance of these features are small. 

\textbf{Putting together.} After checking the separate predictive power of each group, we put multiple groups together to see whether the performance could be further improved. The results are shown in the lower part of Table~\ref{tab:result-session}. We first put the three most useful groups (F10, F11, F12) together with F1, then we find that the result has almost no difference with only using F12. This indicates that F12 is already good enough to represent the battery's discharging pattern, while F10 and F11 only provide little help to it. Then, we put all session features (F5-F18) together with F1 to train the models. The result shows that the RMSE is reduced a little bit, indicating that session features other than F10-F12 could provide some information as a combination although they are not very meaningful individually. Another interesting result is, RF performs better than GBRT and XGB when putting all session features together, which has never happened in other cases. This might because that RF could work better when there are more features.

\subsection{User-specific Feature}

\begin{table}[htbp]
\centering
\caption{Performance of user-specific features. Significant improvements (RMSE $\leq125$, $\tau\geq 0.75$, C-idx $\geq 0.94$) are highlighted.}
\label{tab:result-user}
\begin{tabular}{|l|l|r|r|r|r|}
\hline
\multirow{2}{*}{\textbf{Feature}} & \multirow{2}{*}{\textbf{Metrics}} & \multicolumn{4}{c|}{\textbf{Model}} \\
\cline{3-6}
 & & \multicolumn{1}{c|}{Linear} & \multicolumn{1}{c|}{GBRT} & \multicolumn{1}{c|}{RF} & \multicolumn{1}{c|}{XGB} \\
\hline
\multirow{3}{*}{F1, F19} & RMSE & 135.3 & 132.0 & 137.4 & 131.9 \\
 & Tau & 0.7318 & \textbf{0.7528} & 0.7311 & \textbf{0.7524} \\
 & C-idx & 0.9262 & 0.9311 & 0.9216 & 0.9309 \\
\hline
\multirow{3}{*}{F1, F20} & RMSE & 132.6 & \textbf{121.9} & \textbf{123.9} & \textbf{122.4} \\
 & Tau & 0.7369 & \textbf{0.7680} & \textbf{0.7569} & \textbf{0.7666} \\
 & C-idx & 0.9265 & 0.9359 & 0.9299 & 0.9353 \\
\hline
\multirow{3}{*}{F1, F21} & RMSE & 138.2 & 129.5 & 138.6 & 129.9 \\
 & Tau & 0.7373 & 0.7470 & 0.7348 & 0.7489 \\
 & C-idx & 0.9278 & 0.9317 & 0.9257 & 0.9320 \\
\hline
\multirow{3}{*}{F1, F19-F21} & RMSE & 129.7 & \textbf{118.2} & \textbf{118.9} & \textbf{118.2} \\
 & Tau & 0.7453 & \textbf{0.7710} & \textbf{0.7783} & \textbf{0.7742} \\
 & C-idx & 0.9309 & 0.9391 & 0.9379 & \textbf{0.9400} \\

\hline
\end{tabular}
\end{table}

The rationale of why session features can work is that users' usage has an inertia, so we can infer their usage pattern in the future based on their behavior in the past. This success motivates us to think wider on how to use historical information. Till now, the features we use are limited within the same session. However, users' may have their own usage habit. Therefore, users' usage in the future can be reflected by how they use the device in previous sessions to some degree. As the final step of our experiment, we try three groups of features that describe user's identity and usage in previous sessions: user index (F19), session history (F20), and screen history (F21). Their detailed explanations are:

\begin{itemize}
    \item \textbf{User index (F19).} Each user is allocated one distinct number to identify her identity. Since we have 51 users in the data set, the user index is from 1 to 51. Since different users may have different usage habits, this value can help us better to distinguish users' usage pattern.
    \item \textbf{Session history (F20).} The stats of the average battery consuming rate of the user's previous sessions, both mean and medium are considered here. We add this group of features under the hypothesis that user's usage behavior has an inertia, so that their battery consuming rate before can have a correlation with the current consuming rate. Considering that user's usage may not be same in different weekdays and hours, we add some other filters: only taking into account sessions that begin in the same hour with the current one, in the same weekday with the current one, or both. Finally, we have 4*2=8 features in this group.
    \item \textbf{Screen history (F21).} F21 is the stats of the average percentage of screen-on time of the user's previous sessions. The calculating method is similar to F20.
\end{itemize}

F20 and F21 reflect users' previous usage from two aspects. The former one describes the usage's effect (battery consuming), while the latter one describes the usage state itself. We call these three groups \textbf{``user-specific feature''}. Although they are simple so far, we believe it is still a meaningful initial step to look into these three feature groups. 

The results are shown in Table~\ref{tab:result-user}. F19 and F20 can provide an improvement, though they are not competitive with F10-F12. This shows that user's battery consuming patterns have a weak correlation among different sessions. On the other hand, F21 is not helpful. That may be because this feature is not representative enough. When putting F19-F21 together, the results are keeping better, showing that user-specific features are useful.

\subsection{Putting All Together}

\begin{table}[htbp]
\centering
\caption{Performance if we put features together. Best performances under each metric are highlighted.}
\label{tab:result-together}
\begin{tabular}{|l|l|r|r|r|r|}
\hline
\multirow{2}{*}{\textbf{Feature}} & \multirow{2}{*}{\textbf{Metrics}} & \multicolumn{4}{c|}{\textbf{Model}} \\
\cline{3-6}
 & & \multicolumn{1}{c|}{Linear} & \multicolumn{1}{c|}{GBRT} & \multicolumn{1}{c|}{RF} & \multicolumn{1}{c|}{XGB} \\
\hline
\multirow{3}{*}{F1-F4} & RMSE & 140.2 & 134.3 & 138.3 & 133.6 \\
 & Tau & 0.7223 & 0.7443 & 0.7323 & 0.7452 \\
 & C-idx & 0.9226 & 0.9296 & 0.9239 & 0.9298 \\
\hline
\multirow{3}{*}{F1-F18} & RMSE & HUGE & 108.7 & 114.9 & 106.8 \\
 & Tau & 0.7301 & 0.8088 & 0.8208 & 0.8124 \\
 & C-idx & 0.9282 & 0.9510 & 0.9525 & 0.9520 \\
\hline
\multirow{3}{*}{F1-F21} & RMSE & HUGE & \textbf{106.7} & 110.5 & 107.8 \\
 & Tau & 0.7268 & 0.8142 & \textbf{0.8266} & 0.8160 \\
 & C-idx & 0.9275 & 0.9524 & \textbf{0.9547} & 0.9528 \\

\hline
\end{tabular}
\end{table}

After introducing all the features, we put all groups together to see how good the performance is. As listed in Table~\ref{tab:result-together}, the final result is consistent with what we have discussed above. Query-time features are not very informative. However, it has a big improvement if we add session features into the model. In addition, user-specific features make a slight improvement to the model. Finally, our model reduces RMSE from 140.4 to 106.7, raises $\tau$ from 0.7436 to 0.8266, and raises C-idx from 0.9260 to 0.9547. 

\subsection{Discussion}

\subsubsection{Performance analysis}

According to the aforementioned results, query-time features hardly make improvements to the prediction accuracy, but session features can substantially enhance the prediction performance. In addition, user-specific features can also provide some predictive power, though it is not as powerful as session features. To validate whether the results are reliable, we make a group of statistic tests to examine whether the improvements are exactly significant. Here we use the Bootstrap test and the shifted method introduced by Smucker et al.~\cite{smucker2007comparison}. To be more specific, when trying to test the significance of the difference between the performance of two models, we randomly draw cases with replacement from the original test set until we have drawn the same number of cases as in the experiment. In this case, we can calculate the performance of the two compared models under each metric on the new test set, and then get the difference between the two compared models. We repeat this process for 10,000 times to generate the bootstrap distribution, and then use the shifted method to test whether the improvement is significant based on the bootstrap distribution. The test results are presented in Table~\ref{tab:stat-individual}. We select the best results of every single feature group under each metric, and test whether it is significantly better than the baseline. From the table, we can see that query-time features can not significantly outperform the baseline in terms of the Tau and the C-idx, but it can produce significant improvements in terms of the RMSE. Session features and user-specific features can always bring significant improvements. Finally, we make the same tests on the combination of all features, and the results indicate that the best gained performance is also significantly better than the baseline.

\begin{table}[htbp]
\centering
\caption{Statistic test of each individual feature group. Each result is compared with the baseline (use only F1 as the feature). *** indicates significance at 0.01 level. Query-time feature could not significantly outperform the baseline under Tau and C-idx, but it can generate significant improvement under RMSE. Session feature and user-specific features could always generate significant improvements. The results generated by the combination of entire features are also significantly better than the baseline.}
\label{tab:stat-individual}
\begin{tabular}{|l|r|r|r|}
\hline
\textbf{Feature} & \multicolumn{1}{c|}{\textbf{Best RMSE}} & \multicolumn{1}{c|}{\textbf{Best Tau}} & \multicolumn{1}{c|}{\textbf{Best C-idx}} \\
\hline
F1-F4       & 133.6*** & 0.7452 & 0.9298 \\
F1, F5-F18  & 107.5*** & 0.8226*** & 0.9532*** \\
F1, F19-F21 & 118.2*** & 0.7783*** & 0.9400*** \\
F1-F21 & 106.7*** & 0.8266*** & 0.9547*** \\
\hline
\end{tabular}
\end{table}

The experiment results show that the battery consuming pattern itself is very effective to make an accurate prediction. We are very interested in why this simple and intuitive group of features could work so well. Ideally, the battery life could be perfectly predicted by only using the battery consuming rate in the past if the consuming rate is constant. Hence, the battery consuming pattern is more powerful when the battery consuming rate is more stable. Otherwise, the battery consuming pattern will lose its ability if the battery consuming rate during a session has a large variance. Let us suppose the following scenario: a user A goes to work every day and she always goes back home at 6 p.m.. She does not use her smartphone very much while in office, but she prefers to watch videos on her way home. Hence, the battery consuming rate under two states (at work \& on the way home) are extremely different. If the query occurs just at the transit between the two states, it would be very inaccurate to make the prediction only based on consuming pattern. However, this should be the situation that user-specific features can help. If we can use the information of A's previous sessions which at similar time, we will notice that her battery consuming rate is going to rapidly increase, which can greatly help the prediction.

Having this hypothesis, we are eager to conduct a new experiment to check whether it is correct. The general idea of this experiment is splitting the training set into several groups according to their battery consuming stability, and to see different features effect on each group. So the first critical step is how to define the stability of a session. We take the following method: for each session, we calculate how much time does it take to consume each 1\% battery. For example, if a session begins at 80\% and ends at 10\%, then we calculate a list $(tb_{80}, tb_{79}, ..., tb_{11})$ to represent the time it takes to consume the battery from 80\% to 79\%, from 79\% to 78\%, and so on. Then, we calculate the variance of the mentioned list. Apparently, if a session's battery consuming rate is stable, the differences among $\vec{tb}$ should be small, and therefore the variance of the list would be small. We use this variance to demonstrate the stability of the sessions.

\begin{figure}
	\centering
	\begin{center}
		\includegraphics[width=0.5\textwidth]{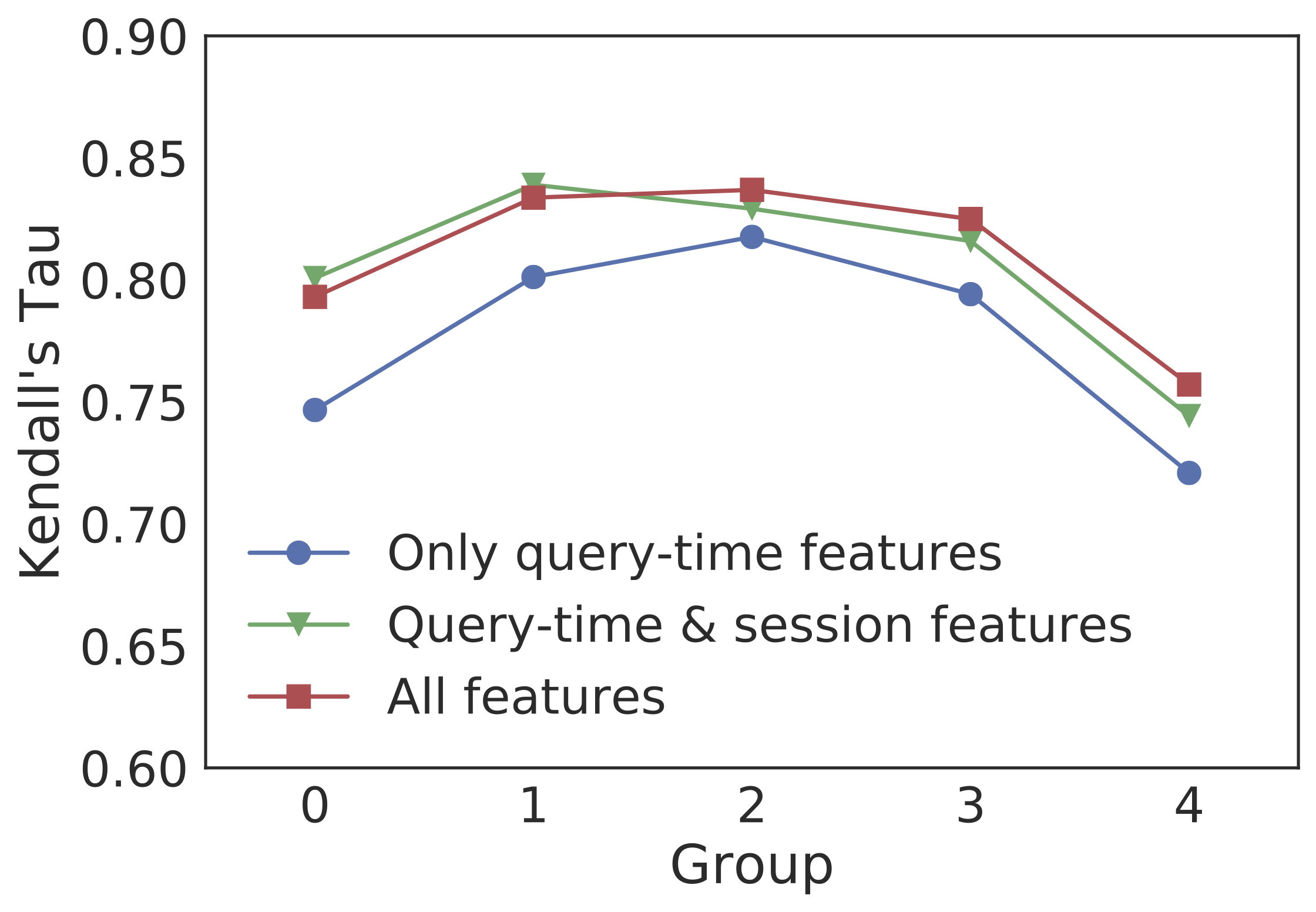}
		\caption[7.5pt]{The $\tau$ of the 5-fold experiment. Each line shows a feature set's results on each session group. The blue line is when only using query-time features, the green line is when adding session features in, and the red line is when using all features. Group 0 has the lowest variance while group 4 has the highest variance.}
		\label{fig:discussion}
	\end{center}
\end{figure}

After we get all sessions' variance, we split the sessions in the training set into 5 groups according to their variance, and then make a 5-fold experiment. Each time, we take out one session group as the validation set, and take the other 4 groups as the training set, to see how well can the regression models perform. We first train the model only based on query-time features, then add session features in, and finally involve all features. The result of $\tau$ is illustrated in Figure~\ref{fig:discussion}. Group 0 has the lowest variances while group 5 has the highest variances. Generally, the result is better if we use more features, which is consistent with our previous discussion. If we make a deeper comparison between the green line and the red line, something interesting happens. When adding user-specific features into the model, the performance on group 0 and group 1 decrease. This indicates that when the session's battery consuming rate is stable, the battery usage pattern is powerful, and the user-specific features even make the performance worse. However, if the variance is large, the user-specific features become useful, and it is getting more effective with the increasing of the variance. The improvements of group 2-4 are 0.008, 0.009, and 0.013. Our hypothesis is verified by the experiment result.

\subsubsection{Survival Analysis Models}

We have carefully considered the problem of censored data and that is why we used C-index as the evaluation metric. Ideally, a best performing prediction model should have been a machine learning model that directly handles censored data. There are such models in survival analysis, such as Cox regression~\cite{cox1992regression}, CoxBoost~\cite{cox1992regression}, Tobit~\cite{tobin1958estimation}, etc. However, during our experiments, we have tried various kinds of survival analysis models, but none of them have a good performance. Most of them simply failed to handle the data at our scale, especially when dealing with more than 1,000 features. For those which did not fail, the prediction accuracy is relatively poor. This is perhaps not surprising, as survival analysis for large-scale and high-dimensional data is an exploratory research direction and the current methods are not robust or scalable yet. Since the goal of our first study is rather demonstrating the feasibility of predicting battery lifetime than maximizing the prediction accuracy, we decided to use the well established and scalable machine learning algorithms. We do note that the prediction performance may be still underestimated and plan to systematically explore large-scale survival analysis in our future work.

\section{Limitations}
As a prediction task, our approach inherently could have some limitations that can affect the generalizability of our results. The first potential threat is that the data set is collected from a single device model, i.e., Samsumg Galaxy S5. Some features are device-specific and may not be available on other device models. However, such type of features account for very little of all features, e.g., only some specific sensors. As a result, we believe that our prediction model itself can be generalized to most device models, as most of the features used in our model are can be collected from standard Android system, and can be even obtained on any similar devices (such as iPhone). In addition, our model follows the standard machine learning pipeline, so it could be applied to any data sets. 

Another possible limitation is that the users in our data set are mainly from a controlled user group from Israel. Therefore, some of our findings may not always hold for users from other regions. However, we regard that our findings are still meaningful because most features (especially features that have strong predicting power) rely on the information that can be collected at  system-level rather than subjective user-specific behaviors. 
\section{Conclusion}

In this paper, we have proposed a novel method for predicting the battery life of smartphones, which is built upon a longitudinal, comprehensive, and fine-grained real-world data set and state-of-the-art machine learning models. Our method is the first one which uses concordance index in survival analysis to address data missing problem in this context. This metric is practical to many real-world scenarios as the data missing problem is quite pervasive in data mining tasks.

The results of the experiments indicate that the remaining battery life of a smartphone could be predicted through a comprehensive set of features and well designed methods. Our method successfully reduces the prediction errors by 33 minutes on average, and it produces significant improvements in both Kendall's Tau and concordance index. the improvement of the accuracy by 33 minutes is quite meaningful in practice. 30 minutes is sufficient for users to adjust their smartphone usage. For example, they can use this period of time to find a power resource, or make a few important phone calls or other critical actions before the phone goes offline. According to the  literature~\cite{hu2017roaming}, the average length of a usage session of smartphones is less than 7 minutes. 33 minutes can be a gamechanger in reality, during which users can change the priority of the tasks and finish the most important ones.

Among all the features, we find that system status at the query time are almost not helpful on top of the current battery level. However, if we take into account the past system status in the same session, the prediction accuracy is largely enhanced. Particularly, the battery's discharging history is the most predictive feature, indicating that the battery's discharging behavior itself is the most important factor. User-specific features also provide additional predictive power, indicating that there is room for personalized prediction of battery life. We find that the system status are more powerful when the discharging rate is stable. Otherwise, user-specific features will be more informative. The overall performance of tree-based models is observed to be better than the linear regression, and thus validates the hypothesis that tree-based models can outperform the linear regression model. It indicates that the battery lifetime does have some non-linear relationships between system status and users' usage behavior.

These findings could be directly useful for related stakeholders as well as provide useful insights to other device models. Since the majority of the features used by our model (battery information, time, sensors, apps' running log, etc.) are quite general and can also be found on any modern smartphones, it is easy to generalize our model to any other smartphone, or even other related contexts such as wearable devices and electric vehicles. Although we cannot guarantee that our model can always produce the same results on different device models, the results presented in this paper can still help to build new prediction models on other device models by techniques such as transfer learning. In addition. It is because that our model follows standard machine learning pipelines, and anyone who has the similar data sets can entirely re-train the model to obtain new results for their own contexts.

In the future, we plan to improve our work from two aspects. On the one hand, we would like to include more data dimensions, such as more sensors and more detailed app status. On the other hand, we will explore more advanced machine-learning models to achieve better prediction results. A particular plan is to explore models for survival analysis (e.g., Cox regression model, random survival trees, Tobit, and so on).

\bibliographystyle{unsrt}
\bibliography{battery}

\begin{thebibliography}{10}

\bibitem{balasubramanian2009energy}
Niranjan Balasubramanian, Aruna Balasubramanian, and Arun Venkataramani.
\newblock Energy consumption in mobile phones: a measurement study and
  implications for network applications.
\newblock In {\em Proceedings of the 9th ACM SIGCOMM conference on Internet
  measurement conference}, pages 280--293, 2009.

\bibitem{puustinen2011effect}
Ismo Puustinen and Jukka~K Nurminen.
\newblock The effect of unwanted internet traffic on cellular phone energy
  consumption.
\newblock In {\em New Technologies, Mobility and Security (NTMS), 2011 4th IFIP
  International Conference on}, pages 1--5, 2011.

\bibitem{rosen2015revisiting}
Sanae Rosen, Ashkan Nikravesh, Yihua Guo, Z~Morley Mao, Feng Qian, and
  Subhabrata Sen.
\newblock Revisiting network energy efficiency of mobile apps: Performance in
  the wild.
\newblock In {\em Proceedings of the 2015 ACM conference on internet
  measurement conference}, pages 339--345, 2015.

\bibitem{shen2015enhancing}
Haichen Shen, Aruna Balasubramanian, Anthony LaMarca, and David Wetherall.
\newblock Enhancing mobile apps to use sensor hubs without programmer effort.
\newblock In {\em Proceedings of the 2015 ACM International Joint Conference on
  Pervasive and Ubiquitous Computing}, pages 227--238, 2015.

\bibitem{he2015optimizing}
Songtao He, Yunxin Liu, and Hucheng Zhou.
\newblock Optimizing smartphone power consumption through dynamic resolution
  scaling.
\newblock In {\em Proceedings of the 21st Annual International Conference on
  Mobile Computing and Networking}, pages 27--39, 2015.

\bibitem{chen2015smartphone}
Xiaomeng Chen, Abhilash Jindal, Ning Ding, Yu~Charlie Hu, Maruti Gupta, and
  Rath Vannithamby.
\newblock Smartphone background activities in the wild: Origin, energy drain,
  and optimization.
\newblock In {\em Proceedings of the 21st Annual International Conference on
  Mobile Computing and Networking}, pages 40--52, 2015.

\bibitem{draa2015application}
Ismat~Chaib Draa, Jamel Tayeb, Smail Niar, and Emmanuelle Grislin.
\newblock Application sequence prediction for energy consumption reduction in
  mobile systems.
\newblock In {\em Computer and Information Technology; Ubiquitous Computing and
  Communications; Dependable, Autonomic and Secure Computing; Pervasive
  Intelligence and Computing (CIT/IUCC/DASC/PICOM), 2015 IEEE International
  Conference on}, pages 23--30, 2015.

\bibitem{li2015characterizing}
Huoran Li, Xuan Lu, Xuanzhe Liu, Tao Xie, Kaigui Bian, Felix~Xiaozhu Lin,
  Qiaozhu Mei, and Feng Feng.
\newblock Characterizing smartphone usage patterns from millions of android
  users.
\newblock In {\em Proceedings of the 2015 ACM Conference on Internet
  Measurement Conference}, pages 459--472, 2015.

\bibitem{falaki2010diversity}
Hossein Falaki, Ratul Mahajan, Srikanth Kandula, Dimitrios Lymberopoulos,
  Ramesh Govindan, and Deborah Estrin.
\newblock Diversity in smartphone usage.
\newblock In {\em Proceedings of the 8th international conference on Mobile
  systems, applications, and services}, pages 179--194, 2010.

\bibitem{mirsky2016sherlock}
Yisroel Mirsky, Asaf Shabtai, Lior Rokach, Bracha Shapira, and Yuval Elovici.
\newblock Sherlock vs moriarty: A smartphone dataset for cybersecurity
  research.
\newblock In {\em Proceedings of the 2016 ACM Workshop on Artificial
  Intelligence and Security}, pages 1--12, 2016.

\bibitem{shye2009into}
Alex Shye, Benjamin Scholbrock, and Gokhan Memik.
\newblock Into the wild: studying real user activity patterns to guide power
  optimizations for mobile architectures.
\newblock In {\em Microarchitecture, 2009. MICRO-42. 42nd Annual IEEE/ACM
  International Symposium on}, pages 168--178, 2009.

\bibitem{dong2011self}
Mian Dong and Lin Zhong.
\newblock Self-constructive high-rate system energy modeling for
  battery-powered mobile systems.
\newblock In {\em Proceedings of the 9th international conference on Mobile
  systems, applications, and services}, pages 335--348, 2011.

\bibitem{zhang2010accurate}
Lide Zhang, Birjodh Tiwana, Robert~P Dick, Zhiyun Qian, Z~Morley Mao, Zhaoguang
  Wang, and Lei Yang.
\newblock Accurate online power estimation and automatic battery behavior based
  power model generation for smartphones.
\newblock In {\em Hardware/Software Codesign and System Synthesis (CODES+
  ISSS), 2010 IEEE/ACM/IFIP International Conference on}, pages 105--114, 2010.

\bibitem{min2015sandra}
Chulhong Min, Chungkuk Yoo, Inseok Hwang, Seungwoo Kang, Youngki Lee, Seungchul
  Lee, Pillsoon Park, Changhun Lee, Seungpyo Choi, and Junehwa Song.
\newblock Sandra helps you learn: the more you walk, the more battery your
  phone drains.
\newblock In {\em Proceedings of the 2015 ACM international joint conference on
  Pervasive and ubiquitous computing}, pages 421--432, 2015.

\bibitem{min2015powerforecaster}
Chulhong Min, Youngki Lee, Chungkuk Yoo, Seungwoo Kang, Sangwon Choi, Pillsoon
  Park, Inseok Hwang, Younghyun Ju, Seungpyo Choi, and Junehwa Song.
\newblock Powerforecaster: Predicting smartphone power impact of continuous
  sensing applications at pre-installation time.
\newblock In {\em Proceedings of the 13th ACM Conference on Embedded Networked
  Sensor Systems}, pages 31--44, 2015.

\bibitem{dong2014rethink}
Mian Dong, Tian Lan, and Lin Zhong.
\newblock Rethink energy accounting with cooperative game theory.
\newblock In {\em Proceedings of the 20th annual international conference on
  Mobile computing and networking}, pages 531--542, 2014.

\bibitem{pathak2011fine}
Abhinav Pathak, Y~Charlie Hu, Ming Zhang, Paramvir Bahl, and Yi-Min Wang.
\newblock Fine-grained power modeling for smartphones using system call
  tracing.
\newblock In {\em Proceedings of the sixth conference on Computer systems},
  pages 153--168, 2011.

\bibitem{mittal2012empowering}
Radhika Mittal, Aman Kansal, and Ranveer Chandra.
\newblock Empowering developers to estimate app energy consumption.
\newblock In {\em Proceedings of the 18th annual international conference on
  Mobile computing and networking}, pages 317--328, 2012.

\bibitem{manweiler2011avoiding}
Justin Manweiler and Romit Roy~Choudhury.
\newblock Avoiding the rush hours: Wifi energy management via traffic
  isolation.
\newblock In {\em Proceedings of the 9th international conference on Mobile
  systems, applications, and services}, pages 253--266, 2011.

\bibitem{li2014making}
Ding Li, Angelica~Huyen Tran, and William~GJ Halfond.
\newblock Making web applications more energy efficient for oled smartphones.
\newblock In {\em Proceedings of the 36th International Conference on Software
  Engineering}, pages 527--538, 2014.

\bibitem{zhao2011system}
Xia Zhao, Yao Guo, Qing Feng, and Xiangqun Chen.
\newblock A system context-aware approach for battery lifetime prediction in
  smart phones.
\newblock In {\em Proceedings of the 2011 ACM Symposium on Applied Computing},
  pages 641--646, 2011.

\bibitem{kang2011personalized}
Joon-Myung Kang, Sin-seok Seo, and James Won-Ki Hong.
\newblock Personalized battery lifetime prediction for mobile devices based on
  usage patterns.
\newblock {\em Journal of Computing Science and Engineering}, 5(4):338--345,
  2011.

\bibitem{kim2016accurate}
Dongwon Kim, Yohan Chon, Wonwoo Jung, Yungeun Kim, and Hojung Cha.
\newblock Accurate prediction of available battery time for mobile
  applications.
\newblock {\em ACM Transactions on Embedded Computing Systems (TECS)},
  15(3):48, 2016.

\bibitem{aharony2011social}
Nadav Aharony, Wei Pan, Cory Ip, Inas Khayal, and Alex Pentland.
\newblock Social fmri: Investigating and shaping social mechanisms in the real
  world.
\newblock {\em Pervasive and Mobile Computing}, 7(6):643--659, 2011.

\bibitem{armstrong2001evaluating}
J~Scott Armstrong.
\newblock Evaluating forecasting methods.
\newblock In {\em Principles of forecasting}, pages 443--472. 2001.

\bibitem{hyndman2006another}
Rob~J Hyndman and Anne~B Koehler.
\newblock Another look at measures of forecast accuracy.
\newblock {\em International journal of forecasting}, 22(4):679--688, 2006.

\bibitem{manning2008introduction}
Christopher~D Manning, Prabhakar Raghavan, Hinrich Sch{\"u}tze, et~al.
\newblock {\em Introduction to information retrieval}, volume~1.
\newblock 2008.

\bibitem{harrell1982evaluating}
Frank~E Harrell, Robert~M Califf, David~B Pryor, Kerry~L Lee, and Robert~A
  Rosati.
\newblock Evaluating the yield of medical tests.
\newblock {\em Jama}, 247(18):2543--2546, 1982.

\bibitem{breiman2001random}
Leo Breiman.
\newblock Random forests.
\newblock {\em Machine learning}, 45(1):5--32, 2001.

\bibitem{friedman2002stochastic}
Jerome~H Friedman.
\newblock Stochastic gradient boosting.
\newblock {\em Computational Statistics \& Data Analysis}, 38(4):367--378,
  2002.

\bibitem{friedman2001greedy}
Jerome~H Friedman.
\newblock Greedy function approximation: a gradient boosting machine.
\newblock {\em Annals of statistics}, pages 1189--1232, 2001.

\bibitem{chen2016xgboost}
Tianqi Chen and Carlos Guestrin.
\newblock Xgboost: A scalable tree boosting system.
\newblock In {\em Proceedings of the 22nd acm sigkdd international conference
  on knowledge discovery and data mining}, pages 785--794, 2016.

\bibitem{pedregosa2011scikit}
Fabian Pedregosa, Ga{\"e}l Varoquaux, Alexandre Gramfort, Vincent Michel,
  Bertrand Thirion, Olivier Grisel, Mathieu Blondel, Peter Prettenhofer, Ron
  Weiss, Vincent Dubourg, et~al.
\newblock Scikit-learn: Machine learning in python.
\newblock {\em Journal of Machine Learning Research}, 12(Oct):2825--2830, 2011.

\bibitem{smucker2007comparison}
Mark~D Smucker, James Allan, and Ben Carterette.
\newblock A comparison of statistical significance tests for information
  retrieval evaluation.
\newblock In {\em Proceedings of the sixteenth ACM conference on Conference on
  information and knowledge management}, pages 623--632, 2007.

\bibitem{cox1992regression}
David~R Cox.
\newblock Regression models and life-tables.
\newblock In {\em Breakthroughs in statistics}. 1992.

\bibitem{tobin1958estimation}
James Tobin.
\newblock Estimation of relationships for limited dependent variables.
\newblock {\em Econometrica: journal of the Econometric Society}, 1958.

\bibitem{hu2017roaming}
Ziniu Hu, Yun Ma, Qiaozhu Mei, and Jian Tang.
\newblock Roaming across the castle tunnels: an empirical study of inter-app
  navigation behaviors of android users.
\newblock {\em arXiv preprint arXiv:1706.08274}, 2017.

\end{thebibliography}

\end{document}